\documentclass[journal,10pt]{IEEEtran}

\usepackage{cite}
\usepackage{graphicx}
\usepackage{amsmath}
\usepackage{array}
\usepackage{mdwmath}
\usepackage{amssymb}
\usepackage{mdwtab}
\usepackage{stfloats}
\usepackage{amsmath,amsthm}
\usepackage{threeparttable}
\usepackage{color}
\usepackage{url}
\usepackage{algpseudocode}
\usepackage{algorithm}
\usepackage{multicol}
\usepackage{amssymb}
\usepackage{epstopdf}
\usepackage{balance}
\usepackage[caption=false,font=footnotesize,labelfont=rm,textfont=rm]{subfig}

\algrenewcommand{\algorithmicrequire}{\textbf{Input:}}
\algrenewcommand{\algorithmicensure}{\textbf{Output:}}

\newtheorem{proposition}{Proposition}

\newtheorem{lemma}{Lemma}

\usepackage{algorithm}
\usepackage{algpseudocode}

\begin{document}

\title{Dual-IRS Aided Near-/Hybrid-Field SWIPT: \\ 
	 Passive Beamforming and Independent  \\ Antenna Power Splitting Design}

\author{Chaoying Huang, Wen Chen, Qingqing Wu, Xusheng Zhu, Zhendong Li, Ying Wang, and Jinhong Yuan
 \vspace{-0.5cm}
 
\thanks{
This work was supported in part by the National Key R\&D Program of China under Grant 2023YFB2905000, the Science and Technology Commission of Shanghai Municipality under Grant 24DP1500500, the National Natural Science Foundation of China (NSFC) under Grants 62531015 and U25A20399, and the Shanghai Jiao Tong University 2030 Initiative. 
\textit{(Corresponding author: Wen Chen.)}
}
\thanks{C. Huang, W. Chen, Q. Wu, and X. Zhu are with the Department of Electronic Engineering, Shanghai Jiao Tong University, Shanghai 200240, China (e-mail: chaoyinghuang@sjtu.edu.cn; wenchen@sjtu.edu.cn; qingqingwu@sjtu.edu.cn; xushengzhu@sjtu.edu.cn).
}
\thanks{Z. Li is with the School of Information and Communication Engineering, Xi’an Jiaotong University, Xi’an 710049, China (e-mail: lizhendong@xjtu.edu.cn).
}
\thanks{Y. Wang is with the State Key Laboratory of Networking and Switching Technology, Beijing University of Posts and Telecommunications, Beijing 100876, China (e-mail: wangying@bupt.edu.cn).
}
\thanks{J. Yuan is with the School of Electrical Engineering and Telecommunications, University of New South Wales, Sydney, NSW 2052, Australia (e-mail: j.yuan@unsw.edu.au).
}
}

\markboth{}
{}

\maketitle

\begin{abstract}
%Intelligent reflecting surface (IRS) has emerged as a cost-effective solution to enhance simultaneous wireless information and power transfer (SWIPT) performance via passive signal reflection. 
%Most existing studies on intelligent reflecting surface (IRS)-aided simultaneous wireless information and power transfer (SWIPT) employed simplified far-field modeling and applied equal power splitting (PS) to receiving antennas. However, due to practical deployment requirements, AP/user is often positioned in the near-field of IRS, rendering the far-field model ineffective. Moreover, non-negligible interference in practice  imposes considerable pressure on the PS between information and energy.
This paper proposes a novel dual-intelligent reflecting surface (IRS) aided interference-limited simultaneous wireless information and power transfer (SWIPT) system with independent power splitting (PS), where each receiving antenna applies different PS factors to offer an advantageous trade-off between the useful information and harvested energy. We separately establish the near- and hybrid-field channel models for IRS-reflected links to evaluate the performance gain more precisely and practically. Specifically, we formulate an optimization problem of maximizing the harvested power by jointly optimizing dual-IRS phase shifts, independent PS ratio, and receive beamforming vector in both near- and hybrid-field cases.
%The formulated problem is non-convex, making it challenging to solve directly. 
In the near-field case, the alternating optimization algorithm is proposed to solve the non-convex problem by applying the Lagrange duality method and the difference-of-convex (DC) programming. In the hybrid-field case, we first present an interesting result that the AP-IRS-user channel gains are invariant to the phase shifts of dual-IRS, which allows the optimization problem to be transformed into a convex one. Then, we derive the asymptotic performance of the combined channel gains in closed-form and analyze the characteristics of the dual-IRS. Numerical results validate our analysis and indicate the performance gains of the proposed scheme that dual-IRS-aided SWIPT with independent PS over other benchmark schemes.

\end{abstract}

\section{Introduction}
With the rapid evolution toward the sixth-generation (6G) networks, the Internet-of-Things (IoT) is expected to achieve unprecedented levels of connectivity, intelligence, and automation \cite{you2021visio,tat2021opp,qin2024anten}. 
%6G-enabled IoT will empower massive-scale applications, such as smart cities, autonomous transportation, industrial automation, and digital twins, requiring ultra-reliable, low-latency communication (URLLC) and sustainable energy solutions \cite{khan2021spec}, \cite{leta2022arti}. 
Simultaneous wireless information and power transfer (SWIPT), which enables IoT devices to harvest energy from radio frequency (RF) signals while receiving information, is promising to solve the energy scarcity problem for energy-constrained wireless nodes \cite{zhang2013mimo}. 
%Applying SWIPT, mobile nodes can achieve self-sustainability, which brings great convenience to establish green wireless networks \cite{cler2019fun}, \cite{zeng2017com}. 
As one of the operating schemes for SWIPT systems, the power splitting (PS) scheme aims to split the received signal into two different power streams, one for the information decoder (ID) and the other for the energy harvester (EH) \cite{shi2014joint}. Based on the PS scheme, the receiver can adaptively adjust the power splitting ratio according to the real-time signal-to-noise ratio (SNR) and energy requirements, achieving a low-latency system response \cite{Lu2020coo}, \cite{Sun2022Adap}. 

In most existing SWIPT works, an equal PS scheme is applied to different receiving antennas \cite{zhu2015wire}, \cite{zong2016opt}, assuming a uniform power trade-off between information decoding and energy harvesting, since the equal PS can enable near-optimal performance for multi-antenna systems in interference-free scenarios \cite{zhang2013mimo}. However, users' quality-of-service (QoS) is usually degraded by interference nodes (INs) in practical wireless networks, and the receiving antennas may encounter diverse channel conditions and varying energy requirements. The conventional equal PS scheme fails to accommodate such dynamic fluctuations adequately. With the aim of overcoming the above difficulties, independent power control for SWIPT networks has been investigated in some literature \cite{xiang2018energy}, \cite{tang2020joi}. In \cite{xiang2018energy}, independent PS ratios, which were applied to different antennas of amplify-and-forward (AF) relay, were jointly optimized with the sources and relay precoding matrices to maximize the harvested energy in two-way AF relay networks. Compared to the equal PS protocol, employing different PS ratios for users leads to a preferable equivalent-sum-rate performance in a multi-user SWIPT system \cite{tang2020joi}. Although independent PS has shown great potential in optimizing the trade-off between information decoding and energy harvesting, severe propagation loss between the access point (AP) and receiver would decrease the QoS–energy region sharply as the distance increases, thus greatly limiting the performance of the SWIPT system \cite{li2022uav}.

Recently, intelligent reflecting surface (IRS) has emerged as a cost-efficient technology to reconstruct the wireless channel and provide additional links for blocked direct links \cite{chen2023statmi,zhang2024omni,zhu2024robust}. Based on the advantages of the IRS, the momentum of the IRS-assisted SWIPT networks has been stimulated \cite{li2022allo,sha2024Fractional,li2022beam}. In \cite{sha2024Fractional}, a common PS ratio at the end-user and transmit power were jointly optimized by using a practical phase-dependent amplitude model for each IRS element reflectivity. \cite{li2022beam} proposed a novel network framework of IRS-assisted SWIPT NOMA systems with independent PS control for different users. Nevertheless, the question of how to exploit optimal independent PS for different receiving antennas in the IRS-aided SWIPT system remains largely open. 
In addition, it is necessary to inject new vitality into the IRS deployment design to meet the challenges of the high-quality connectivity and cost-effective deployment \cite{huang2025elem}. Motivated by the benefits of partitioning a single large IRS into multiple smaller IRSs \cite{kang2022irs}, a dual-IRS architecture was investigated in \cite{han2022double}, which demonstrated superior performance to a single IRS with an equal total number of reflecting elements. From the perspective of implementation, installing two IRSs is more cost- and maintenance-efficient than deploying multiple IRSs simultaneously, making it particularly suitable for small- to medium-scale communication scenarios (e.g., indoor coverage or smart venues) and early experimental validation. Several recent studies have confirmed that dual-IRS architectures can significantly improve network performance \cite{feng2023dualdis}, \cite{yan2025dopp}. For instance, it has been shown that dual-IRS can be placed on two building facades facing different directions to provide angular diversity and coverage enhancement \cite{yan2025dopp}. Additionally, \cite{wu2021intelligent} observed that placing the IRS near the AP or user yields the largest SNR. Given the highly dynamic nature of user terminal locations, deploying IRS near the user presents significant implementation challenges, some works proposed integrating the IRS with the AP to achieve higher channel rank and passive beamforming gain \cite{zheng2023simu}, \cite{huang2023bs}.
Whereas, with the distance between the IRS and the AP/user shrinking and the aperture of IRS increasing to improve the system performance \cite{wu2021intelligent}, the electromagnetic radiation characteristics of the IRS will gradually evolve from the far-field radiation with planar wavefront to the near-field radiation with spherical wavefront.

In near-field scenarios, the conventional far-field uniform plane wave (UPW) assumption may become inapplicable, and several new channel characteristics with the more generic non-uniform spherical wave (NUSW) propagation need to be considered \cite{zhou2015spher,fried2019local,son2021primer}, such as variations of signal amplitude and phase across array elements \cite{feng2021wier}, and spatial channel non-stationarity \cite{mar2020anten}. 
According to the new channel characteristics of near-field scenarios, the conventional far-field passive beamforming design for IRS-aided SWIPT systems \cite{li2022allo}, \cite{li2022beam}, \cite{zhao2022irs}, \cite{tang2023ener} may be invalid, as the near-field passive beamforming of IRS needs to be adjusted in the additional distance domain to guarantee the performance of IRS-aided SWIPT systems. 
Meanwhile, the increased sensitivity of channel modeling to antenna spacing in the near field \cite{lu2022mimonear} would lead to heterogeneous signal distributions across receiving antennas in IRS-aided SWIPT systems, where specific antennas are located in stronger signal regions and are thus more suitable for information decoding, while others are better suited for energy harvesting, thereby motivating the study of independent antenna PS optimization.
Furthermore, although one of the two-hop transmissions via the IRS (i.e., AP-IRS link and IRS-user link) would evolve into a near-field transmission \cite{wu2021intelligent}, the other hop may be in either far-field or near-field, so there may be a hybrid of far and near fields in IRS-aided SWIPT systems. In this case, the simple near/far field propagation model becomes insufficient, and the passive beamforming design of IRS becomes more challenging due to the requirement to jointly consider the channel conditions of the AP-IRS and the IRS-user links.

Motivated by the superior performance of dual-IRS over a single IRS, as well as their cost- and maintenance-efficiency for deployment, we are encouraged to adopt a dual-IRS architecture and integrate the dual-IRS with the AP, i.e., AP is always located in the near field of the dual-IRS, thereby exploiting the improved spatial rank and beamforming benefits while ensuring high-quality connectivity.
Considering user mobility, the user may reside in either the near field or the far field.
To this end, we focus on investigating a dual-IRS aided SWIPT network with independent antenna PS in this paper for two channel cases: the near-field channel and the hybrid-field channel. Specifically, under the constraint of meeting the user QoS requirement, we maximize the harvested energy by jointly optimizing the independent PS ratio, receiving beamforming vector, and dual-IRS phase shifts for two different channel models, respectively. 
The main contributions are summarized as follows:
\begin{itemize}
\item We propose a dual-IRS aided interference-limited SWIPT system with independent PS, where each antenna at the receiver can apply different PS factors to adjust the ratio between information reception and energy harvesting flexibly. To accurately characterize system performance, we establish both a near-field model (for AP-IRS and IRS-user links) and a hybrid-field model (near-field for AP-IRS links and far-field for IRS-user links).   
To optimize system performance in different scenarios, we formulate the harvested energy maximization problem for joint optimization of the independent PS ratio, receive beamforming vector, and dual-IRS phase shifts in the near-field and hybrid-field cases, respectively. Due to the strong coupling among optimization variables, the problem is non-trivial. For the near-field case, we decompose the non-convex problem into two subproblems and alternately optimize them by applying Lagrange duality method and the difference-of-convex (DC) programming.
\item Moreover, based on the hybrid-field channel modeling, we derive the combined channel gains of the AP-IRS1/IRS2-user links that provide valuable insights. In particular, our results reveal the independence of combined channel gains from the dual-IRS phase shifts, transforming the optimization problem into a convex one. Additionally, we derive closed-form expressions for the combined channel gains and analyze their asymptotic performance to further investigate the characteristics of the dual-IRS. These results show that regardless of the dimension in which the dual-IRS expands, the combined channel gains tend to approach constant values. When both dimensions of the IRS in its plane are sufficiently large, the dual-IRS become like double mirrors that reflect half of the AP transmit power in their front half-space reflection area, respectively.
\item Through numerical simulation, we validate the performance of the proposed scheme of the dual-IRS aided SWIPT with independent PS for both the near-field model and the hybrid-field model. It is shown that the proposed scheme outperforms other benchmark schemes, i.e., adopting independent antenna PS for the user and dual-IRS for AP assistance can significantly increase the harvested power. Besides, it can be obtained that irrespective of whether the user is in the near- or far-field of the dual IRS, the closer the dual IRS is to the AP, the better the SWIPT performance. Meanwhile, we demonstrate the necessity of proper dual-IRS channel modeling for accurately characterizing the considered system performance.
\end{itemize}

The remainder of this paper is organized as follows. In Section \uppercase\expandafter{\romannumeral2}, we elaborate on the channel models for near-field and hybrid-field, and propose the signal model and optimization problem formulation for the dual-IRS aided SWIPT system. Section \uppercase\expandafter{\romannumeral3} presents the optimization algorithms proposed for the formulated problems in the near-field and hybrid-field models, respectively, and analyzes the characteristics of the dual-IRS in the hybrid-field model. In Section \uppercase\expandafter{\romannumeral4}, simulation results are provided to assess the effectiveness of the proposed scheme in both channel models and to validate our analytical findings. Finally, the conclusion is given in Section \uppercase\expandafter{\romannumeral5}.

\textit{Notations}: Scalars are denoted by lowercase letters, vectors by bold lowercase letters, and matrices by bold uppercase letters. $\mathbb{C} ^{M\times N}$ denotes the space of $M\times N$ complex-valued matrices. For a complex-valued scalar $x$, $\left | x \right | $ denotes the absolute value. For a complex-valued vector $\boldsymbol{x}$, $\left \| \boldsymbol{x} \right \| $ denotes the Euclidean norm, $\left \| \boldsymbol{x} \right \| _{\ell  } $ denotes its $\ell $-norm, and $\mathrm{diag } (\boldsymbol{x} )$ denotes a diagonal matrix with the elements of $\boldsymbol{x}$ on its main diagonal. For a square matrix $\mathbf{X}$, $\mathbf{X}^{-1} $ and $\mathrm{Tr} (\mathbf{X} )$ denote its inverse and trace, respectively, while $\mathbf{X} \succeq 0  $ represents that X is positive semi-definite. For a general matrix $\mathbf{S}$, $\mathrm{rank} \left (\mathbf{S}  \right )$, $\mathbf{S} ^{H} $, and $ \left [  \mathbf{S} \right ]  _{m,n} $ denote its rank, conjugate transpose, and $\left ( m,n \right ) $-th entry, respectively. A circularly symmetric complex Gaussian (CSCG) random vector with zero mean and covariance matrix $\mathbf{C} $ is denoted as $\mathcal{CN}\left ( 0,\mathbf{C}  \right )$, and $\sim $ stands for “distributed as”.

\section{System Model and Problem Formulation}
\begin{figure}[t]
	\centering
	\includegraphics[width=8cm]{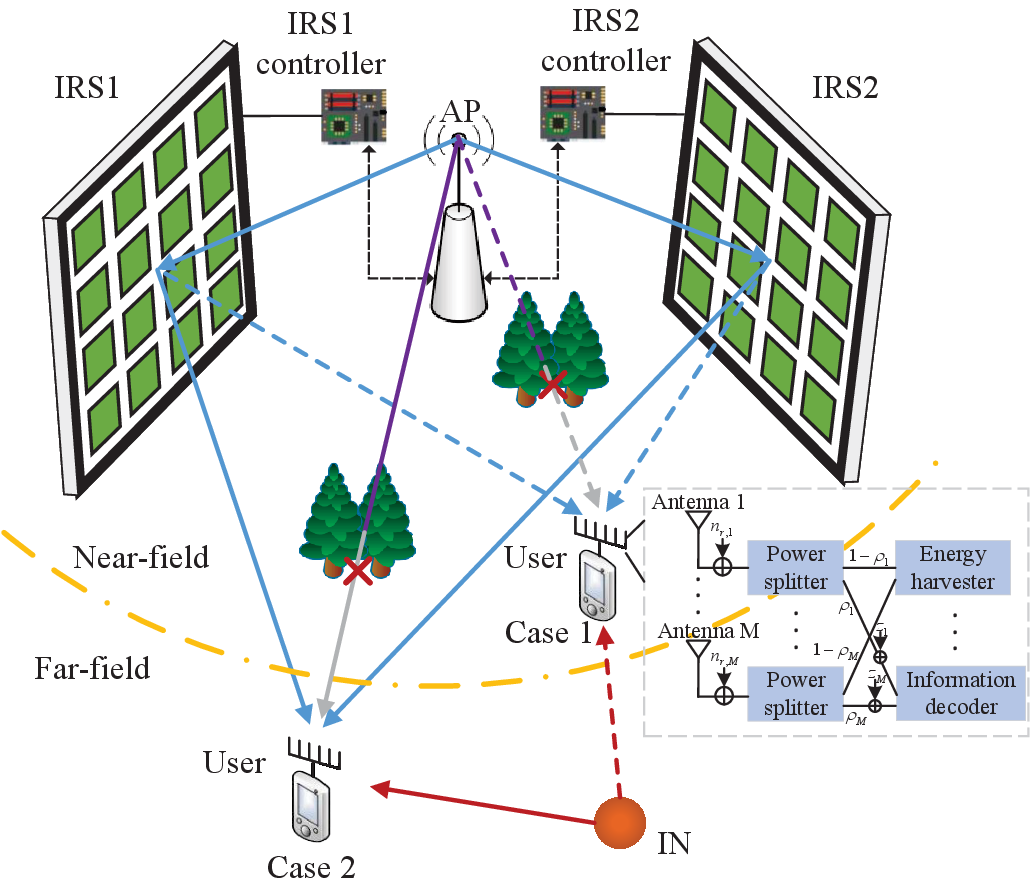}\\
	\caption{\small{The near-field dual-IRS aided interference-limited SIMO SWIPT system.}}\label{system}
	\vspace{-0.2cm}
\end{figure}

As shown in Fig. \ref{system}, we consider a dual-IRS-aided SWIPT system, where a single-antenna AP adjacent to two auxiliary IRSs serves a multi-antenna user\footnote{The results presented in this work can be extended to multi-user scenarios, provided that orthogonal frequency bands are allocated to different users, as in IRS-aided orthogonal frequency-division multiple access (OFDMA) systems \cite{yang2020ofdm}. With inter-user interference taken into account, new optimization methods (e.g., fractional programming \cite{shen2018frac}) should be employed in future work.}, suffering a dominant interference from IN. In the considered system, the dual-IRS integrated with the AP performs the beamforming functionality, making a single-antenna AP sufficient for system operation. The direct path between the AP and the user is assumed to be blocked due to the existence of propagation obstacles. The user is equipped with an $M$-element uniform linear array (ULA) of antennas and the IN is equipped with a single antenna. The two IRSs are defined as IRS1 and IRS2 respectively, consisting of a uniform planar array (UPA), which are assumed to be placed on the x-z plane. IRS1 and IRS2 are deployed with $N_a = N_{x_a} \times N_{z_a}$ and $N_b = N_{x_b} \times N_{z_b}$ passive reflecting elements, where $N_{x_a}(N_{x_b})$ and $N_{z_a}(N_{z_b})$ denote the number of elements along the $x$-axis and $z$-axis, respectively. Integrating with an intelligent controller, the IRS can dynamically adjust the phase shift of each reflecting element based on the channel state information (CSI). To characterize the maximum performance gain of IRS-aided SWIPT in this paper, we assume that the CSI of all relevant links is perfectly known at the AP. As the IRS is a passive reflecting device, we adopt a time-division duplexing (TDD) protocol for uplink and downlink transmissions and exploit channel reciprocity to acquire downlink CSI based on the uplink training \cite{li2022beam}.

%\footnote{\textcolor{blue}{In the considered system, the dual-IRS integrated with the AP performs the beamforming functionality, making a single-antenna AP sufficient for system operation. This work can be extended to the multi-antenna AP case by jointly optimizing the AP’s active beamforming, the IRSs’ passive beamforming, the independent antenna PS, and the receive beamforming vector.}}

%\footnote{\textcolor{blue}{This work can be readily extended to multi-IRS systems by assigning each IRS an independent reflection matrix, where the block coordinate descent (BCD) method may be applied to update these matrices iteratively or in parallel to ensure computational efficiency and convergence \cite{Wu2020Qos}; Under the hybrid-field condition, similar principles apply by modeling each IRS–transceiver link separately and optimizing independent PS factors via CVX or the interior-point method.}}

Two different channel models are considered for the dual-IRS-aided SWIPT system, as shown in Fig. \ref{system}. Specifically, as the dual-IRS are always closed to the AP, these two cases are illustrated respectively as (a) Near-field model for both AP-IRS (i.e., AP-IRS1 and AP-IRS2) and IRS-user (i.e., IRS1-user and IRS2-user) links; (b) Hybrid-field model that includes near-field channels for AP-IRS links and far-field channels for IRS-user links. We assume that the system operates at a frequency of $f_0$, so the wavelength of the signal is $\lambda _{0} = c f_{0} $. Denote the element spacing along the x- and z-axis as $\varepsilon $, which is shown in Fig. \ref{3d}. For the IRS1 with a UPA, the array aperture is calculated as $D_a=\sqrt{\left [ \left ( N_{x_a}-1 \right ) \varepsilon \right ]^{2} + \left [ \left ( N_{z_a}-1 \right ) \varepsilon \right ]^{2} } $, which is similar to IRS2. The far-field and near-field regions of IRS1 and IRS2 are divided by the Rayleigh distances $R_a=\frac{2D_{a}^{2}  }{\lambda _{0} } $ and $R_b=\frac{2D_{b}^{2}  }{\lambda _{0} } $, respectively.

\subsection{Channel Model}
To accurately characterize the considered dual-IRS-aided SWIPT system, lay the foundation for precise system performance optimization, and extend the generality of the considered dual-IRS-aided SWIPT framework, we first provide mathematical descriptions of the near- and hybrid-field models.

\begin{figure}[t]
	\centering
	\includegraphics[width=7cm]{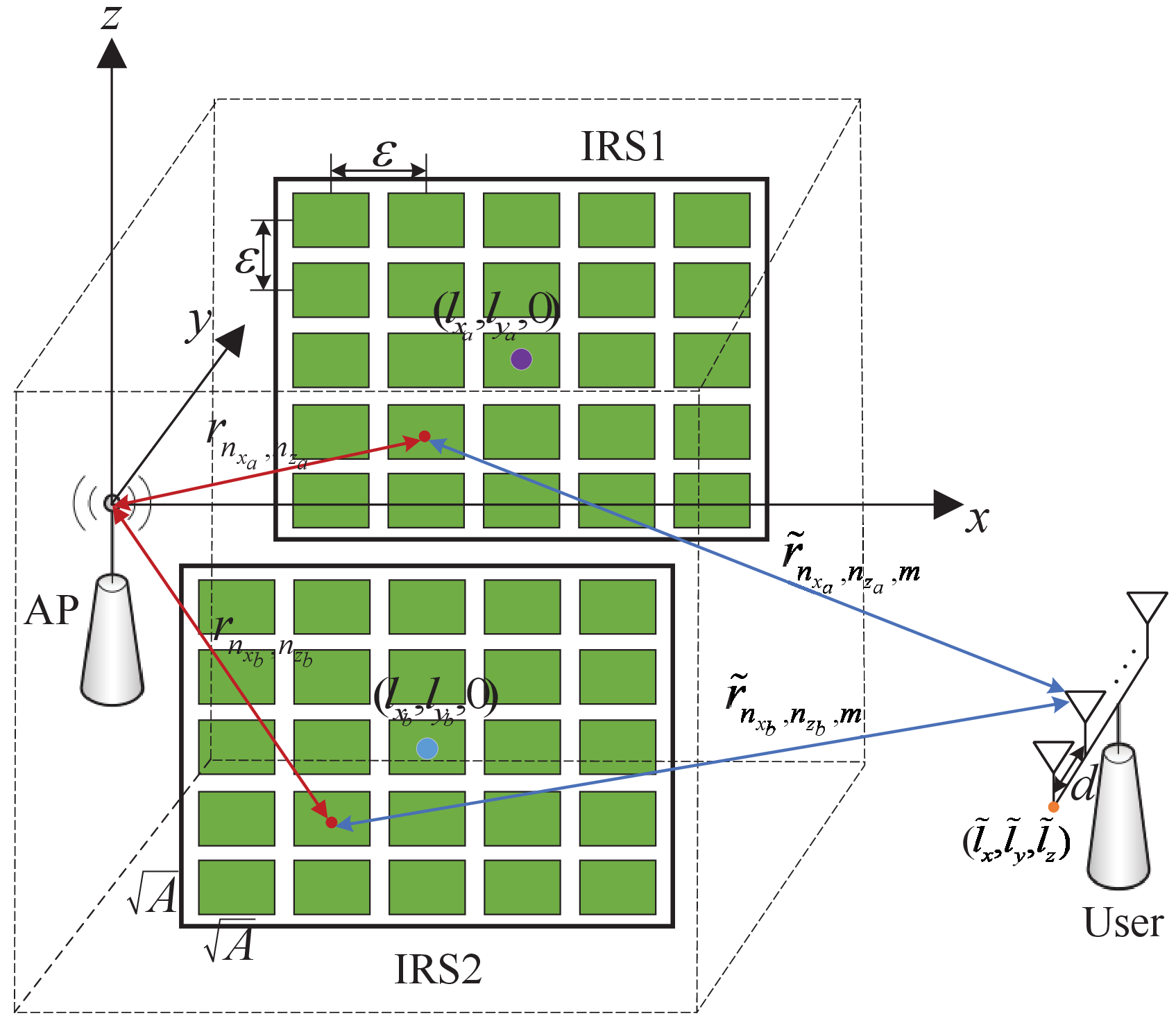}\\
	\caption{\small{The near-field geometry relationship in the 3D Cartesian coordinate system.}}\label{3d}
	\vspace{-0.2cm}
\end{figure}

For AP-IRS links, the near-field channel model we establish is described below. For ease of exposition, we assume that the AP antenna is placed at the origin of the three-dimensional (3D) Cartesian coordinate system illustrated in Fig. \ref{3d}, i.e., $ \mathrm{p} _\mathrm{A} =\left ( 0,0,0 \right ) $. Denoted the physical size of each reflecting element as $\sqrt{A} \times \sqrt{A}$ with $\sqrt{A} \le \varepsilon $. For notational convenience, we assume that the number of elements along the x- and z-axis is odd, and the array centers of IRS1 and IRS2 are placed at $\left ( l_{x_a},l_{y_a},0   \right ) $ and $\left ( l_{x_b},l_{y_b},0   \right ) $, respectively. Denote the central locations of the $\left ( n_{x},n_{z}  \right ) $-th element of IRS1 and IRS2 as $ \mathrm{p}_{n_{x_a},n_{z_a} }=\left [ l_{x_a}+ n_{x_a}\varepsilon , l_{y_a},n_{z_a}\varepsilon  \right ] $ and $ \mathrm{p}_{n_{x_b},n_{z_b} }=\left [ l_{x_b}+ n_{x_b}\varepsilon , l_{y_b},n_{z_b}\varepsilon  \right ] $ respectively, where $ n_{x_a} = 0,\pm 1,...,\pm \left ( N_{x_a}-1  \right )/2 $, $ n_{z_a} = 0,\pm 1,...,\pm \left ( N_{z_a}-1  \right )/2 $, $ n_{x_b} = 0,\pm 1,...,\pm \left ( N_{x_b}-1  \right )/2 $ and $ n_{z_b} = 0,\pm 1,...,\pm \left ( N_{z_b}-1  \right )/2 $. As such, the distances between the AP and the centers of the $\left ( n_{x},n_{z}  \right ) $-th element of IRS1 and IRS2 can be given by 
\begin{align}
r_{n_{x_a},n_{z_a}} \! \!  & = \!  \left \| \mathrm{p} _\mathrm{A}  \! -  \! \mathrm{p}_{n_{x_a},n_{z_a} } \right \|_{2} \!  =  \! \sqrt{\left ( l_{x_a}+ n_{x_a}\varepsilon \right )^{2} \!  + \! l_{y_a}^{2} \! + \! \left ( n_{z_a}\varepsilon \right )^{2}    } \notag  \\ 
&= l_{x_a}\sqrt{1+\bar{r}_a^2+2n_{x_a}\xi _{a}+\left ( n_{x_a}^2+ n_{z_a}^2\right ) \xi _{a}^2 }   ,
\end{align}
\begin{align}
r_{n_{x_b},n_{z_b}} \! \!  & = \!  \left \| \mathrm{p} _\mathrm{A}  \! -  \! \mathrm{p}_{n_{x_b},n_{z_b} } \right \|_{2} \!  =  \! \sqrt{\left ( l_{x_b}+ n_{x_b}\varepsilon \right )^{2} \!  + \! l_{y_b}^{2} \! + \! \left ( n_{z_b}\varepsilon \right )^{2}    } \notag  \\ 
&= l_{x_b}\sqrt{1+\bar{r}_b^2+2n_{x_b}\xi _{b}+\left ( n_{x_b}^2+ n_{z_b}^2\right ) \xi _{b}^2 }   ,
\end{align}
where $\bar{r}_a=\frac{l_{y_a}}{l_{x_a}}$, $\bar{r}_b=\frac{l_{y_b}}{l_{x_b}}$, $\xi _{a}=\frac{\varepsilon }{l_{x_a}} $ and $\xi _{b}=\frac{\varepsilon }{l_{x_b}} $. Note that the element spacing is typical on the order of sub-wavelength at the signal carrier frequency; thus we have $\xi _{a},\xi _{b}\ll 1$ in practice.

Based on 3D channel modeling and the projected aperture non-uniform spherical wave (PNUSW) model, which accounts for the spherical wavefront and the variations in signal amplitude and projected aperture across array elements \cite{lu2022comm}, the channel power gain between the AP antenna and the $\left ( n_{x_a},n_{z_a}  \right ) $-th element of IRS1 can be derived as
\begin{align} \label{am1}
q  _{ n_{x_a},n_{z_a} } & = \frac{1}{4\pi \left \| \mathrm{p} _\mathrm{A}   -   \mathrm{p}_{n_{x_a},n_{z_a} } \right \|_{2}^{2}  }
\underbrace{A\frac{\left ( \mathrm{p} _\mathrm{A}  \! -  \mathrm{p}_{n_{x_a},n_{z_a} } \right )^{T} \mathbf{u} _a }{\!  \left \| \mathrm{p} _\mathrm{A}  \! -   \mathrm{p}_{n_{x_a},n_{z_a} } \right \|_{2} \! }}_{\text{projected aperture}} \notag \\
&=\frac{Al_{y_a}}{4\pi l_{x_a}^{3} \left [ 1+\bar{r}_a^2+2n_{x_a}\xi _{a}+\left ( n_{x_a}^2+ n_{z_a}^2\right ) \xi _{a}^2    \right ] ^{\frac{3}{2} }  }   ,
\end{align}
where $ \mathbf{u} _a $ denotes the normal vector of each IRS1 element placed
on the x-z plane., i.e., $\mathbf{u} _a=(0,-1,0)$. According to the channel gain model in (\ref{am1}), the array response vector, denoted as $\mathbf{h}_a \in \mathbb{C} ^{N_a\times 1} $, is then formed by the following elements
\begin{equation}
h_{n_{x_a},n_{z_a}}  =\sqrt{q  _{ n_{x_a},n_{z_a} }} e^{-j\frac{2\pi }{\lambda _{0} }r_{n_{x_a},n_{z_a}} }, \ \forall n_{x_a},n_{z_a} .
\end{equation}

The channel power gain between the AP antenna and the $\left ( n_{x},n_{z}  \right ) $-th element of IRS2 is similar to the model in (\ref{am1}), i.e., 
\begin{align} \label{am2}
q  _{ n_{x_b},n_{z_b} }\! =\frac{-Al_{y_b}}{4\pi l_{x_b}^{3} \left [ 1+\bar{r}_b^2+2n_{x_b}\xi _{b}+\left ( n_{x_b}^2+ n_{z_b}^2\right ) \xi _{b}^2    \right ] ^{\frac{3}{2} }  }   ,
\end{align}
Similarly, the elements of channel response vector $\mathbf{h}_b \in \mathbb{C} ^{N_b\times 1} $ can be given by 
\begin{equation}
h_{n_{x_b},n_{z_b}}  =\sqrt{q  _{ n_{x_b},n_{z_b} }} e^{-j\frac{2\pi }{\lambda _{0} }r_{n_{x_b},n_{z_b}} }, \ \forall n_{x_b},n_{z_b}  .
\end{equation}

For IRS-user links, when the distance between the IRS and user is less than the Rayleigh distance, the channels between the user and IRS are built as the near-field model, which is illustrated as follows. Denote the array spacing of the ULA of antennas mounted at the user as $d$. Assuming that the linear array is placed along the direction of the y-axis, in which starting point is located at $\left ( \tilde{l}_{x},\tilde{l}_{y},\tilde{l}_{z}   \right ) $, and thus the location of the $m$-th antenna of the user is $\mathrm{p}_{m}  =\left ( \tilde{l}_{x},\tilde{l}_{y}+md,\tilde{l}_{z}   \right ) $, where $m= 1,2,...,M$. Accordingly, the distance between the $m$-th antenna of the user and the centers of the $\left ( n_{x_a},n_{z_a}  \right ) $-th element of IRS1 can be given by 
\begin{align}
&\tilde{r}_{n_{x_a},n_{z_a},m} \!   =  \left \|   \mathrm{p}_{n_{x_a},n_{z_a} } -\mathrm{p}_{m} \right \|_{2} \!   \notag  \\ 
& \!=\sqrt{ \!\left ( \!\tilde{l}_{x}-l_{x_a} \!- \! n_{x_a}\varepsilon \! \right )^{2}  \! + \!\left (   \tilde{l}_{y} \!+\!md -l_{y_a} \! \right )^{2}   \!+ \!\left ( \tilde{l}_{z}-n_{z_a}\varepsilon \! \right )^{2}  }   ,
\end{align} 
Similar to Eq. (\ref{am1}), the channel power gain between the $m$-th antenna of the user and the $\left ( n_{x_a},n_{z_a}  \right ) $-th element of IRS1 can be characterized as
\begin{align} \label{am3}
&\tilde{q }  _{ n_{x_a},n_{z_a},m } \notag \\
&=\frac{A \left ( l_{y_a} \!- \! \tilde{l}_{y} \!-\!md  \right ) }{4 \pi \left [  \!\left ( \!\tilde{l}_{x}\!-\!l_{x_a} \!\!- \! n_{x_a}\varepsilon \! \right )^{2}\! \! \! + \!\!\left (   \tilde{l}_{y} \!+\!md -l_{y_a} \! \right )^{2}  \! \!+ \!\!\left ( \tilde{l}_{z}-n_{z_a}\varepsilon \! \right )^{2}   \right ]^\frac{3}{2}  \! } .
\end{align}

Accordingly, the channel response vector of the IRS1-user link, denoted as $\mathbf{G}_a \in \mathbb{C} ^{M\times N_a} $, is then formed by the following elements
\begin{equation} \label{channel}
g^a_{n_{x_a}\!,n_{z_a}\!,m} \!\! =\!\!\sqrt{\tilde{q }  _{ n_{x_a},n_{z_a},m }} e^{-j\frac{2\pi }{\lambda _{0} }\tilde{r}_{n_{x_a},n_{z_a},m}  },  \forall n_{x_a},n_{z_a},m  ,
\end{equation}
The elements of the channel response vector of IRS2-user link $\mathbf{G}_b \in \mathbb{C} ^{M\times N_b}$ are similar to (\ref{channel}), which are omitted for brevity.

Further denote the phase shifts introduced by the $(n_{x_a},n_{z_a})$-th IRS1 element and the $(n_{x_b},n_{z_b})$-th IRS2 element as $\theta _{n_{x_a},n_{z_a}} $ and $\theta _{n_{x_b},n_{z_b}} $, respectively. Moreover, $\mathbf{\Theta } _{a}\in  \mathbb{C} ^{N_a\times N_a } $ and $\mathbf{\Theta } _{b}\in  \mathbb{C} ^{N_b\times N_b } $ are diagonal matrices, with the diagonal entries given by $e^{j\theta _{n_{x_a},n_{z_a}} } $ and $e^{j\theta _{n_{x_b},n_{z_b}} } $, respectively. Then, the total combined channel of the near-field model mentioned above can be expressed as 
\begin{equation}
\mathbf{g}\left ( \boldsymbol{l } \right )   =\mathbf{G} _{a}{\mathbf{\Theta}}  _{a}\mathbf{h} _{a}+\mathbf{G} _{b}{\mathbf{\Theta}}  _{b}\mathbf{h} _{b},
\end{equation}
where $\boldsymbol{l }$ consists of $\left (  l_{x_a},l_{y_a} ,l_{x_b},l_{y_b}, \tilde{l}_{x},\tilde{l}_{y},\tilde{l}_{z} \right )$ representing the coordinates of the dual-IRS and user, and $\mathbf{g}\left ( \boldsymbol{l } \right )   =\left [ g_{1}\left ( \boldsymbol{l } \right ), g_{2}\left ( \boldsymbol{l } \right ),...,g_{M}\left ( \boldsymbol{l } \right ) \right ] ^{T}$. 

On the other hand, for IRS-user links, when the distance between the IRS1/IRS2 and the user is larger than the Rayleigh distance, the channels between the user and IRS1/IRS2, i.e., $\mathbf{G} _{a}$ and $\mathbf{G} _{b}$, should be established as the far-field model. Due to the rich scattering environment at the user's side as well as the long distance between the user and the dual-IRS, $\mathbf{G}_{a}$ and $\mathbf{G} _{b}$ are assumed to follow the Rayleigh fading channel model, which will be further discussed in Section \uppercase\expandafter{\romannumeral4}. Since the severe propagation loss, signals reflected by the two IRSs twice or more are negligible. Besides, we consider the distance between the IN and the user to surpass the Rayleigh distance, and thus the channel of IN-user would be established as the far-field model. Furthermore, the IN is considerably far from the dual-IRS, operating in the far-field region relative to them, and the interference signal reflected by the dual-IRS is negligible due to the sufficiently large distance and severe path loss. It is assumed that the channel from IN to the user is statistically known at the user, which is denoted as $\mathbf{f} \in \mathbb{C} ^{M\times 1} $ with $\mathbf{f}  =\left [ f_{1}, f_{2},...,f_{M} \right ] ^{T} $, following the Rayleigh fading channel model.

\subsection{Signal Model}
Assume that the transmit power of AP and IN are $P_{\mathrm{t} }$ and $P_{\mathrm{in} } $, respectively. The received signal at the user is 
\begin{equation}
{\mathbf{y}}  = (\mathbf{G} _{a}{\mathbf{\Theta}}  _{a}\mathbf{h} _{a}+\mathbf{G} _{b}{\mathbf{\Theta}}  _{b}\mathbf{h} _{b} ) \sqrt{P_{\mathrm{t} } }{ {s}} _{1}+\mathbf{f}   \sqrt{P_{\mathrm{in} } }{s} _{2}+\mathbf{n} _{r},
\end{equation} 
where $s_1$ and $s_2$ are the normalized transmitted signals at the AP and IN, respectively, and $\mathbf{n} _{r}\sim \mathcal{CN} (0,\sigma _{r}^{2}\mathbf{I}  )$ is the antenna noise at the user.

We consider that each receiving antenna of the user applies an independent PS scheme to coordinate support for more flexible signal processing. Specifically, the signal received by each antenna is split to the ID and EH by an individual power splitter as shown in Fig. \ref{system}. For the $m$-th receiving antenna, it divides $\rho_m $ ($0\le \rho_m \le 1$) portion of the signal power to the ID, and the remaining $1-\rho_m$ portion of the signal power to the EH. Assuming the independent PS vector is $\boldsymbol{\rho}  =\left [ \rho _{1},\rho _{2},...,\rho _{M}  \right ] ^{T }  $, the total harvested energy by the EH can be given by 
%\footnote{\textcolor{blue}{Since this work focuses on a typical SWIPT system with low radio frequency (RF) input power, where the EH circuit can be closely approximated by a linear model \cite{Morsi2020condi}, we adopt the linear EH model to ensure analytical tractability. Note that for the non-linear model, the alternating optimization (AO) and successive convex approximation (SCA) techniques can be applied to solve optimization problems \cite{hua2022nonl}.}}
\begin{equation}
Q=\eta \sum_{m=1}^{M} (1-\rho _{m} )(P_\mathrm{t}  \left |g_{m} \right |^{2}+P_\mathrm{in}\left |f_{m} \right |^{2}+\sigma_{r}^{2} ),
\end{equation}
where $\eta\in (0,1]$ denotes the energy conversion efficiency at EH. The adoption of the linear EH model aims to gain fundamental system insights and maintain analytical tractability, which is a widely-used simplification in complex system-level optimizations \cite{xiang2018energy,tang2020joi,li2022uav}, \cite{li2022beam}, \cite{tang2023ener}. We note that while practical EH circuits exhibit non-linearities, the use of a realistic non-linear model, like the logistic function, would lead to a highly non-convex and mathematically intractable optimization problem. Such a problem would typically require advanced techniques, such as the introduction of slack variables and the successive convex approximation (SCA) method \cite{hua2022nonl}.
In this paper, we consider the normalized time, then the harvested energy is the harvested power. After information decoding, the signal split to the ID can be expressed as
\begin{align}
{\mathbf{y}} ^{\mathrm{ID} }  =  & \mathbf{w}^{H} \mathbf{\Lambda } ^{\frac{1}{2} } (\mathbf{G} _{a}{\mathbf{\Theta}}  _{a}\mathbf{h} _{a}+ \mathbf{G} _{b}{\mathbf{\Theta}}  _{b}\mathbf{h} _{b}) 
\sqrt{P_\mathrm{t} }{ {s} } _{1} \notag \\
&+ \mathbf{w}^{H}  \mathbf{\Lambda } ^{\frac{1}{2} } \mathbf{f}   \sqrt{P_{\mathrm{in} } }{s} _{2} +\mathbf{w}^{H}  \mathbf{\Lambda } ^{\frac{1}{2} }\mathbf{n} _{r}+ \mathbf{w}^{H}  \mathbf{z},
\end{align}
where $\mathbf{\Lambda }  = \mathrm{diag} (\boldsymbol{\rho} )$, $\mathbf{z} \sim \mathcal{CN} (0,\delta ^{2}\mathbf{I}  )$ is the additional noise introduced by the ID, and $\mathbf{w}\in \mathbb{C} ^{M\times 1} $ represents the receive beamforming vector.

Accordingly, the signal to the interference-plus-noise ratio (SINR) of the user is
\begin{equation}
\mathrm{SINR}  = \frac{P_\mathrm{t} \left | \mathbf{w}^{H}  \mathbf{\Lambda }^{\frac{1}{2} } (\mathbf{G} _{a}{\mathbf{\Theta}}  _{a}\mathbf{h} _{a}+\mathbf{G} _{b}{\mathbf{\Theta}}  _{b}\mathbf{h} _{b} )  \right |^{2}  }
{P_\mathrm{in} \left | \mathbf{w}^{H}  \mathbf{\Lambda } ^{\frac{1}{2} } \mathbf{f}  \right | ^{2}\!\!\!\!+ \!\sigma _{r} ^{2}\left \| \mathbf{w}^{H}  \mathbf{\Lambda } ^{\frac{1}{2} } \right \|  ^{2} \!\!\!\!+\!\delta  ^{2}\left \|\mathbf{w}  \right \| ^{2} } .
\end{equation}

\subsection{Problem Formulation}
In this paper, we aim to maximize the harvested energy by jointly designing the independent PS ratio, the receive beamforming vector, and dual-IRS phase shift matrices. Hence, the problem can be formulated as follows,
\begin{align}
(\text{P1}) \quad &\underset{ \left \{\mathbf{\Theta}  _{a},\mathbf{\Theta}  _{b},\mathbf{w},\boldsymbol{\rho}  \right \}  }{\text{max}} \quad   Q  \notag  \\
\mbox{s.t.}\quad
&  \frac{P_\mathrm{t} \left | \mathbf{w}^{H}  \mathbf{\Lambda }^{\frac{1}{2} } (\mathbf{G} _{a}{\mathbf{\Theta}}  _{a}\mathbf{h} _{a}+\mathbf{G} _{b}{\mathbf{\Theta}}  _{b}\mathbf{h} _{b} )   \right |^{2}  }  
{P_\mathrm{in} \left | \mathbf{w}^{H}  \mathbf{\Lambda } ^{\frac{1}{2} } \mathbf{f}  \right | ^{2}\!\!\!\!+ \!\sigma _{r} ^{2}\left \| \mathbf{w}^{H} 
 \mathbf{\Lambda } ^{\frac{1}{2} } \right \|  ^{2} \!\!\!\!+\!\delta  ^{2}\left \|\mathbf{w}  \right \| ^{2} }\ge \gamma_0  ,\label{SI0}\\
& 0\le \rho _{m}\le 1 ,\  m=1,2,...,M,  \label{splitting factor0}\\
&0\le {\theta }_{n_{x_a},n_{z_a} },{\theta }_{n_{x_b},n_{z_b} }\le 2\pi \label{phase0} ,
\end{align}
where $ n_{x_a} = 0,\pm 1,...,\pm \left ( N_{x_a}-1  \right )/2 $, $ n_{z_a} = 0,\pm 1,...,\pm \\ \left ( N_{z_a}-1  \right )/2 $, $ n_{x_b} = 0,\pm 1,...,\pm \left ( N_{x_b}-1  \right )/2 $ and $ n_{z_b} = 0,\pm 1,...,\pm \left ( N_{z_b}-1  \right )/2 $. Constraint (\ref{SI0}) guarantees the QoS requirement of the user with the SINR threshold $\gamma_0 $. Considering that the user should have non-zero SINR threshold, i.e., $\gamma_0>0$. In addition, the independent PS ratio of each antenna should satisfy constraint (\ref{splitting factor0}). Constraint (\ref{phase0}) is the condition that dual-IRS phase shifts should meet.

%For the near-field and hybrid-field channel models, the propagation characteristics of the IRS-user link are near-field spherical wave and far-field plane wave, respectively. 
In order to ensure the performance of SWIPT, the passive beamforming of dual-IRS and the independent antenna PS ratio demand to be designed flexibly depending on the propagation characteristics of near-/hybrid-field channel. Therefore, the investigation for SWIPT in near-field and hybrid-field will be discussed separately in the following.

%\begin{figure}
%  \centering
%  \includegraphics[width=8cm]{PS.eps}\\
%  \caption{\small{The independent PS receiver architecture.}}\label{ps}
%  \vspace{-0.2cm}
%\end{figure}

\section{SWIPT Optimization for the Near-field Model}
In this section, we concentrate on dual-IRS-aided SWIPT optimization in near-field channels. It can be seen that the problem (P1) is a non-convex optimization problem, since the optimization variables are highly coupled and the phase shifts are expressed in exponential form. Therefore, we divide the non-convex problem into two sub-problems and alternately optimize the receive beamforming vector, independent PS ratio and dual-IRS phase shift matrices by applying the Lagrange duality method and the DC programming algorithm.

\subsection{Receive Beamforming Vector and Independent PS Ratio Design}
Given dual-IRS phase shift matrices, we apply MMSE for the receive beamforming vector at ID, i.e., $\mathbf{w} = \mathbf{S}^{-1} \mathbf{\Lambda }^{\frac{1}{2}} \mathbf{g}\left ( \boldsymbol{l } \right ) $ with $\mathbf{S}  = P_\mathrm{in} \mathbf{\Lambda }^{\frac{1}{2}} \mathbf{f}\mathbf{f}^{H}\mathbf{\Lambda }^{\frac{1}{2}}+\sigma _{r} ^{2}\mathbf{\Lambda }+\delta  ^{2}\mathbf{I}$ \cite{Ver1998Dete}.
%Furthermore, it is assumed that the receiving antenna spacing $d$ is sufficiently large, resulting in uncorrelated spatial channels across antennas. 
Accordingly, the SINR can be re-expressed as 
\begin{align} \label{sinrb}
	\mathrm{SINR} \! =  &\frac{P_\mathrm{t} \mathbf{w}^{H}\!\mathbf{\Lambda } ^{\frac{1}{2}}\mathbf{g}\left ( \boldsymbol{l } \right )\!\mathbf{g}\left ( \boldsymbol{l } \right ) ^{H}\!\!\mathbf{\Lambda }^{\frac{1}{2}}\mathbf{w}}{\mathbf{w}^{H} \mathbf{S }\mathbf{w}}\!=\!P_\mathrm{t} \mathbf{g}\left ( \boldsymbol{l } \right ) ^{H}\!\!\mathbf{\Lambda }^{\frac{1}{2}} \mathbf{S }^{-1}\! \mathbf{\Lambda } ^{\frac{1}{2}}\mathbf{g}\left ( \boldsymbol{l } \right ) .
%	=& \sum_{m=1}^{M} \frac{P_\mathrm{t} \rho _{m}\left | g_{m } \right | ^{2} }{P_\mathrm{in} \rho _{m}\left | f_{m}  \right | ^{2}+ \rho _{m}\sigma _{r}^{2}+\delta  ^{2}} .
\end{align}

Set $\mathbf{\hat{g}  }\left ( \boldsymbol{l } \right )  = \sqrt{P_\mathrm{t} }  \mathbf{g}\left ( \boldsymbol{l } \right )$ and  $\mathbf{\hat{f} } =\sqrt{P_\mathrm{in} }  \mathbf{f}$, we have $\mathrm{SINR} =\mathbf{\hat{g}  }\left ( \boldsymbol{l } \right )^{H}\tilde{\mathbf{S}} ^{-1} \mathbf{\hat{g}  }\left ( \boldsymbol{l } \right )$ with $\tilde{\mathbf{S}} = \mathbf{\hat{f} }\mathbf{\hat{f} }^{H}  +\delta ^{2}\mathbf{\Lambda } ^{-1}+\sigma _{r}^{2}\mathbf{I} $.
Consequently, the problem (P1) can be transformed into:
\begin{align}
(\text{P2})\quad & \underset{ \left \{0\le \rho _{m}\le 1  \right \}  }{\text{max}} \quad   Q  \notag  \\
\mbox{s.t.}\quad
& \mathbf{\hat{g}  }\left ( \boldsymbol{l } \right )^{H}\tilde{\mathbf{S}} ^{-1} \mathbf{\hat{g}  }\left ( \boldsymbol{l } \right )\ge \gamma_0 .\label{SINR}
%&\mathrm{SINR} \ge \gamma_0 .\label{SINR}
\end{align}

Since the SINR of the user is a concave function with respect to the independent PS ratio $\boldsymbol{\rho}$ \cite{lis2016indep}, the problem (P2) is convex and the Slater’s condition holds \cite{Boyd2004cvx}. Thus, the duality gap is zero. By solving its dual problem, we can obtain its optimal solution. The Lagrangian function corresponding to this problem can be given by
\begin{align}
\mathcal{L} = \eta \sum_{m=1}^{M} (\rho _{m} -1) & (P_\mathrm{t}  \left |g_{m}\left ( \boldsymbol{l } \right ) \right |^{2}+P_\mathrm{in}\left |f_{m} \right |^{2}+\sigma_{r}^{2} )\notag \\
& +\lambda (\gamma_0 -\mathbf{\hat{g}  }\left ( \boldsymbol{l } \right )^{H}\tilde{\mathbf{S}} ^{-1} \mathbf{\hat{g}  }\left ( \boldsymbol{l } \right )),
\end{align}
where $\lambda$ is the Lagrangian multiplier. Next, we use the Karush-Kuhn-Tucker (KKT) conditions to investigate the optimal solution of the dual problem. The KKT conditions is given by
\begin{align} \label{KKT1}
&\mathrm{K1} :\nabla_{\rho _{m} } \mathcal{L}\left ( \rho _{m},\lambda  \right )  = 0, \notag \\
&\mathrm{K2} :\lambda\left ( \gamma_0 -\mathbf{\hat{g}  }\left ( \boldsymbol{l } \right )^{H}\tilde{\mathbf{S}} ^{-1} \mathbf{\hat{g}  }\left ( \boldsymbol{l } \right ) \right )  =0 ,\notag \\
&\mathrm{K3} :\lambda \ge 0.
\end{align}

For the K1 condition, it can be expanded to
\begin{align} 
&\nabla_{\rho _{m} } \mathcal{L}\left ( \rho _{m},\lambda  \right ) \notag \\
&\!= \!\eta \left (\!P_\mathrm{t}  \left |g_{m}\!\left ( \boldsymbol{l } \right ) \right |^{2}\!\!+\!\!P_\mathrm{in}\left |f_{m} \right |^{2}\!\!+\!\sigma_{r}^{2}  \right )
\!-\!\lambda \!\frac{\delta ^{2}\left | {\mathbf{\hat{g}  }\left ( \boldsymbol{l } \right )}\!^{H} \! \left [ \tilde{\mathbf{S}}^{-1}  \right ]_{:,m}    \right |^{2}   }{\rho _{m}^2 } \! =\! 0,
\end{align}
By checking the expansion of K1 and condition K3, one can readily obtain that $\lambda>0$. Then, with the optimal $\lambda$, we have $\gamma_0 -\mathbf{\hat{g}  }\left ( \boldsymbol{l } \right )^{H}\tilde{\mathbf{S}} ^{-1} \mathbf{\hat{g}  }\left ( \boldsymbol{l } \right ) =0$. Additionally, the optimal $\rho _{m}$ should satisfy (\ref{KKT1}), which thus can be expressed as
\begin{equation} \label{rou}
\rho _{m} \! =\! \left [ \frac{\delta\sqrt{\lambda }  \left | \mathbf{\hat{g}  }\left ( \boldsymbol{l } \right )^{H}\!\left [ \tilde{\mathbf{S}}^{-1}  \right ]_{:,m}    \right |}{\sqrt{\eta \!\left (P_\mathrm{t}  \left |g_{m}\left ( \boldsymbol{l } \right ) \right |^{2}\!+\!P_\mathrm{in}\left |f_{m} \right |^{2}\!+\!\sigma_{r}^{2}  \right )} }  \!\right  ]_{0}^{1}\!\!, m=1,2,...,M ,
\end{equation}
where $\left [a\right ] _{0}^{1} =\mathrm{min} \left \{ \mathrm{max} \left \{ a,0 \right \} ,1 \right \}$.

It can be observed that equation (\ref{rou}) is not a closed-form expression. As the right-hand-side (RHS) of this equation contains $\rho _{m}$, we employ fixed-point iteration (FPI) to get the optimal $\rho _{m}^{\ast } $. Then, for the Lagrangian multiplier $\lambda$, we apply the sub-gradient method to derive the optimal $\lambda ^{\ast } $. 
%In order to satisfy the condition $\lambda>0$, an appropriate initial point and step size need to be selected.

\subsection{Dual-IRS Phase Shifts Design}
In this subsection, the dual-IRS phase shifts determination algorithm based on the DC programming is proposed. When the received beamforming vector and independent PS ratio are given, the problem (P1) can be rewritten as
\begin{align}
(\text{P3})\ \    &\underset{ \left \{\mathbf{\Theta}  _{a},\mathbf{\Theta}  _{b}  \right \}  }{\text{max}} \quad   Q  \notag  \\
\mbox{s.t.}\quad
&  \frac{P_\mathrm{t} \left | \mathbf{w}^{H}  \mathbf{\Lambda }^{\frac{1}{2} } (\mathbf{G} _{a}{\mathbf{\Theta}}  _{a}\mathbf{h} _{a}+\mathbf{G} _{b}{\mathbf{\Theta}}  _{b}\mathbf{h} _{b} )   \right |^{2}  }
{P_\mathrm{in} \left | \mathbf{w}^{H}  \mathbf{\Lambda } ^{\frac{1}{2} } \mathbf{f}  \right | ^{2}\!\!\!\!+ \!\sigma _{r} ^{2}\left \| \mathbf{w}^{H} 
 \mathbf{\Lambda } ^{\frac{1}{2} } \right \|  ^{2} \!\!\!\!+\!\delta  ^{2}\left \|\mathbf{w}  \right \| ^{2} }\ge \gamma_0  ,\label{SINR3}\\
&0\le {\theta }_{n_{x_a},n_{z_a} },{\theta }_{n_{x_b},n_{z_b} }\le 2\pi  .
\end{align}
where $\mathbf{\Theta } _{a} = \mathrm{diag}  \left ( e^{j\theta _{1,1} },e^{j\theta _{1,2} },...,e^{j\theta _{N_{x_a},N_{z_a} } }  \right )$ and $\mathbf{\Theta } _{b} = \mathrm{diag}  \left ( e^{j\theta _{1,1} },e^{j\theta _{1,2} },...,e^{j\theta _{N_{x_b},N_{z_b}} }   \right )$. 
Obviously, the problem (P3) is non-convex, as the non-concave objective function and non-convex constraint (\ref{SINR3}). Let $\mathbf{u}_a=[ e^{j\theta _{1,1} },e^{j\theta _{1,2} },...,e^{j\theta _{N_{x_a},N_{z_a} } } ]^{T}$ and $\mathbf{u}_b=[ e^{j\theta _{1,1} },e^{j\theta _{1,2} },...,e^{j\theta _{N_{x_a},N_{z_a} } } ]^{T}$. Then the constraint on $\theta _{n_{x_a},n_{z_a} }$ and ${\theta }_{n_{x_b},n_{z_b} }$ are equivalent to $\left | u_{n_{x_a},n_{z_a}}  \right |   = 1$ and $\left | u_{n_{x_b},n_{z_b}}  \right |   = 1$, respectively.
Let $\mathbf{\Phi } _{a} = \mathbf{G} _{a}\mathrm{diag} (\mathbf{h} _{a}) \in \mathbb{C} ^{M\times N_{a} }$ and $\mathbf{\Phi } _{b} = \mathbf{G} _{b}\mathrm{diag} (\mathbf{h} _{b}) \in \mathbb{C} ^{M\times N_{b} }$, then $\left \|\mathbf{G} _{a}{\mathbf{\Theta}}  _{a}\mathbf{h} _{a}+\mathbf{G} _{b}{\mathbf{\Theta}}  _{b}\mathbf{h} _{b} \right \| ^{2}$ can be written as $\left \| \mathbf{\Phi } _{a}\mathbf{u}_{a}+ \mathbf{\Phi } _{b}\mathbf{u}_{b}\right \| ^{2} $. We introduce auxiliary variables as follows,
\begin{equation}
\mathbf{\Omega}  =\begin{bmatrix}\mathbf{\Phi } _{a}^{H}\mathbf{\Phi } _{a} 
  & \mathbf{\Phi } _{a}^{H}\mathbf{\Phi } _{b}\\\mathbf{\Phi } _{b}^{H} \mathbf{\Phi } _{a} 
  & \mathbf{\Phi } _{b}^{H}\mathbf{\Phi } _{b} 
\end{bmatrix}
, \
\mathbf{\bar{u} } =\begin{bmatrix}\mathbf{u} _{a} 
 \\\mathbf{u} _{b} 
\end{bmatrix}
.
\end{equation}

Consequently, we have $\left \| \mathbf{\Phi } _{a}\mathbf{u}_{a}+ \mathbf{\Phi } _{b}\mathbf{u}_{b}\right \| ^{2} = \mathbf{\bar{u} }^{H}\mathbf{\Omega}\mathbf{\bar{u} } $. Note that $\mathbf{\bar{u} }^{H}\mathbf{\Omega} \mathbf{\bar{u} }=\mathrm{Tr} (\mathbf{\Omega} \mathbf{\bar{u} }\mathbf{\bar{u} }^{H})$. Define $\bar{\mathbf{U}}= \mathbf{\bar{u} }\mathbf{\bar{u} }^{H}$, which need to satisfy $\bar{\mathbf{U}}\succeq \mathbf{0} $ and $ \mathrm{rank} (\bar{\mathbf{U}})=1$. 

Moreover, we let $\mathbf{t}_{a}^{H}  \!\! =\!\!\mathbf{w}^{H}  \mathbf{\Lambda } ^{\frac{1}{2} } \mathbf{G} _{a}\mathrm{diag} (\mathbf{h} _{a})$, $\mathbf{t}_{b}^{H}  \! \!=\!\!\mathbf{w}^{H}  \mathbf{\Lambda }^{\frac{1}{2} } \mathbf{G} _{b}\mathrm{diag} (\mathbf{h} _{b})$, $\mathbf{\Upsilon }_{a}   \! \!=\! \! \mathbf{\Lambda }^{\frac{1}{2} } \mathbf{G} _{a}\mathrm{diag} (\mathbf{h} _{a})$ and $\mathbf{\Upsilon}_{b}  \! \! = \! \!\mathbf{\Lambda } ^{\frac{1}{2} } \mathbf{G} _{b}\mathrm{diag} (\mathbf{h} _{b})$. Then, $\left |\mathbf{w}^{H}  \mathbf{\Lambda } ^{\frac{1}{2} } (\mathbf{G} _{a}{\mathbf{\Theta}}  _{a}\mathbf{h} _{a}+\mathbf{G} _{b}{\mathbf{\Theta}}  _{b}\mathbf{h} _{b})\notag   \right |^{2} = \left | \mathbf{t}_{a}^{H}\mathbf{u}_{a} +\mathbf{t}_{b}^{H}\mathbf{u}_{b}\right |^{2} $ and  $\left \|  \mathbf{\Lambda } ^{\frac{1}{2} } (\mathbf{G} _{a}{\mathbf{\Theta}}  _{a}\mathbf{h} _{a}+\mathbf{G} _{b}{\mathbf{\Theta}}  _{b}\mathbf{h} _{b})\notag  \right \|   ^{2} = \left \| \mathbf{\Upsilon }_{a}\mathbf{u}_{a} +\mathbf{\Upsilon }_{b}\mathbf{u}_{b} \right \| ^{2}$. We introduce auxiliary variables as follows,
\begin{equation}
\mathbf{R}  =\begin{bmatrix}\mathbf{t } _{a}\mathbf{t } _{a} ^{H}
  & \mathbf{t} _{a}\mathbf{t} _{b}^{H}\\\mathbf{t } _{b} \mathbf{t } _{a} ^{H}
  & \mathbf{t } _{b}\mathbf{t } _{b} ^{H}
\end{bmatrix}
, 
\mathbf{E}  =\begin{bmatrix}\mathbf{\Upsilon  } _{a}^{H}\mathbf{\Upsilon } _{a} 
  & \mathbf{\Upsilon } _{a}^{H}\mathbf{\Upsilon } _{b}\\\mathbf{\Upsilon } _{b}^{H} \mathbf{\Upsilon } _{a} 
  & \mathbf{\Upsilon } _{b}^{H}\mathbf{\Upsilon } _{b} 
  \end{bmatrix}.
\end{equation}

Subsequently, we have $\left |  \mathbf{t} _{a}^{H}\mathbf{u}_{a}+ \mathbf{t} _{b}^{H}\mathbf{u}_{b}  \right | ^{2} = \mathbf{\bar{u} }^{H}\mathbf{R} \mathbf{\bar{u} }$ and $\left \| \mathbf{\Upsilon }_{a}\mathbf{u}_{a} +\mathbf{\Upsilon }_{b}\mathbf{u}_{b} \right \| ^{2}=\mathbf{\bar{u} }^{H}\mathbf{E} \mathbf{\bar{u} }$. Note that $\mathbf{\bar{u} }^{H}\mathbf{R}\mathbf{\bar{u} }=\mathrm{Tr} (\mathbf{R}\bar{\mathbf{ U} }  )$ and $\mathbf{\bar{u} }^{H}\mathbf{E}\mathbf{\bar{u} }=\mathrm{Tr} (\mathbf{E}\bar{\mathbf{ U} }  )$.
As such, the problem (P3) can be reformulated as 
\begin{align}
(\text{P3.1}) \ \  &  \ \ \underset{  \bar{\mathbf{ U} }   }{\text{max}} \quad   P_{\mathrm{t} }\mathrm{Tr}\left ( \mathbf{\Omega }  \bar{\mathbf{ U} }  \right )  
+P_{\mathrm{in} }\left (  \left \| \mathbf{f}  \right \|^{2}  -\left \|\mathbf{ \Lambda} ^{\frac{1}{2} }\mathbf{f}    \right \| ^{2}  \right )\notag \\
& \quad \quad  +\sigma _{r}^{2}  \left ( M- \left \|\mathbf{ \Lambda} ^{\frac{1}{2} }   \right \| ^{2}\right )  
-\mathrm{Tr}\left ( \mathbf{E}\bar{\mathbf{U}  }  \right )  \notag     \\
\mbox{s.t.}\quad 
& \frac{ P_{\mathrm{t} }\mathrm{Tr} (\mathbf{R} \bar{\mathbf{U}})}{P_\mathrm{in} \left | \mathbf{w}^{H}  \mathbf{\Lambda } ^{\frac{1}{2} } \mathbf{f}  \right | ^{2}\!\!\!\!+ \!\sigma _{r} ^{2}\left \| \mathbf{w}^{H}  \mathbf{\Lambda } ^{\frac{1}{2} } \right \|  ^{2} \!\!\!\!+\!\delta  ^{2}\left \|\mathbf{w}  \right \| ^{2}}\ge \gamma_0, \label{sinr3.1} \\
& \mathbf{\bar{U}}_{n,n} =1,\ n=1,2,...,N, \label{unit3.1} \\
&  \mathrm{rank} (\bar{\mathbf{U}})=1, \label{rank1} \\
& \mathbf{\bar{U}}\succeq 0,  \label{semi3.1}
\end{align}
where $N=N_{a} +N_{b} $, representing the total number of reflecting elements of dual-IRS.

To address problem (P3.1), we apply the DC programming to convert the non-convex rank-one constraint \cite{hua2021rec}. 

\begin{proposition}\label{pro1}
For a positive semi-definite matrix $\mathbf{F} \in \mathbb{C}^{M\times M} $ with $\mathrm{Tr} (\mathbf{F} )> 0$, the rank-one constraint can be equivalently reformulated as the difference of two convex functions, given by
\begin{equation}\label{eq13}
\mathrm{rank} \left ( \mathbf{F} \right )  =1 \Leftrightarrow    \mathrm{Tr} (\mathbf{F} )-\left \| \mathbf{F} \right \| _{2}=0,
\end{equation} 
where $\mathrm{Tr} (\mathbf{F} )=\sum_{m=1}^{M}\sigma _{m}  \left ( \mathbf{F}   \right ) $, $\sigma _{m}  \left ( \mathbf{F}   \right )$ is the $n$-th largest singular value of matrix $\mathbf{F}$ \cite{li2022allo}.
\end{proposition}

Based on {\bf Proposition 1}, problem (P3.1) can be transformed into
\begin{align}
(\text{P3.2}) \  &  \ \ \underset{  \bar{\mathbf{ U} }   }{\text{max}} \quad   P_{\mathrm{t} }\mathrm{Tr}\left ( \mathbf{\Omega }  \bar{\mathbf{ U} }  \right )  
+P_{\mathrm{in} }\left (  \left \| \mathbf{f}  \right \|^{2}  -\left \|\mathbf{ \Lambda} ^{\frac{1}{2} }\mathbf{f}    \right \| ^{2}  \right )\notag \\
& \quad \quad  +\sigma _{r}^{2}  \left ( M- \left \|\mathbf{ \Lambda} ^{\frac{1}{2} }   \right \| ^{2}\right ) 
-\mathrm{Tr}\left ( \mathbf{E}\bar{\mathbf{U}  }  \right )     \\
\mbox{s.t.}\quad  
&  \mathrm{Tr} (\bar{\mathbf{ U} }  )-\left \| \bar{\mathbf{ U} }  \right \| _{2}=0, \label{rank3.2} \\
& (\ref{sinr3.1}),(\ref{unit3.1}),(\ref{semi3.1}).
\end{align}

As the left-hand-side (LHS) of constraint (\ref{rank3.2}) is not affine, problem (P3.2) is still non-convex. In this paper, we incorporate constraint (\ref{rank3.2}) into the objective function as a penalty term, making this problem feasible. Furthermore, to guarantee this penalty term convexity, the concave one, i.e., $-\left \| \bar{\mathbf{ U} }  \right \| _{2}$, needs to be convex. The essence of DC programming lies in converting the original problem into a convex optimization problem by linearizing the concave term. Specifically, the following problem in the $t$-th iteration needs to be solved:
\begin{align}
(\text{P3.3}) \ \  &  \ \ \underset{  \bar{\mathbf{ U} }   }{\text{max}} \quad   f\left ( \bar{\mathbf{ U} }  \right )  - \mu  \left (  \mathrm{Tr} (\bar{\mathbf{ U} }  )-
\left \langle \partial \left \| \bar{\mathbf{ U} } ^{t-1}  \right \| _{2}, \bar{\mathbf{ U} } \right \rangle  \right ) \notag  \\
\mbox{s.t.}\quad  
& (\ref{sinr3.1}),(\ref{unit3.1}),(\ref{semi3.1}),
\end{align}
where $ f\left ( \bar{\mathbf{ U} }  \right )$ is the  objective function of problem (P3.2), $\mu $ is a large constant that serves as a penalty factor, and $\partial \left \| \bar{\mathbf{ U} } ^{t-1}  \right \| _{2}$ is the subgradient of the spectral norm of the solution obtained at the $t-1$ iteration.
%\begin{proposition}\label{pro2}
%Given that $\mathbf{F}$ is positive semi-definite, the subgradient $\partial \left \| \mathbf{F} \right \| _{2}$ of its spectral norm is given by $\mathbf{f} _{1} \mathbf{f} _{1} ^{H} $, where $\mathbf{f} _{1}$ denotes the eigenvector associated with the largest singular value of $\mathbf{F}$ \cite{tao1997ana}.
%\end{proposition}
Until now, the problem (P3) has been transformed into a convex optimization problem, which can be solved using the CVX toolbox \cite{grant2016cvx}. In summary, the overall optimization algorithm is presented in {\bf Algorithm 1}.
%Specifically, we set the initial penalty factor $\mu >0$, convergence threshold $\epsilon>0$, and amplification factor $c>1$, and then solve the convex problem (P3.3). If $\beta \left ( \mathrm{Tr} (\bar{\mathbf{ U} }  )-\left \| \bar{\mathbf{ U} }  \right \| _{2} \right ) < \epsilon$ is satisfied, the output $\bar{\mathbf{ U} }$ is the optimal solution on the $t$-th iteration; Otherwise, let $ \mu = c\mu$, and re-solve problem (P3.3) until convergence. 

Notably, although our study primarily focuses on a dual-distributed-IRS aided SWIPT system, the proposed algorithm can be readily extended to general multi-distributed-IRS architectures. Specifically, the original non-convex optimization problem can still be decomposed into two sub-problems as described above, which are optimized alternately. Assume that there are $L$ IRSs, and denote $\mathbf{\Theta } _{l}\in  \mathbb{C} ^{N_l\times N_l } $ the phase shift matrix of the $l$-th IRS. The total combined channel can then be expressed as $\mathbf{g} = \sum_{l=1}^{L} \mathbf{G} _{l}{\mathbf{\Theta}}  _{l}\mathbf{h} _{l}$. With the increase in the number of phase shift matrices, the dimension of the associated auxiliary variables shall be expanded, i.e., 
\begin{equation}
	\mathbf{\Omega}  =
	\begin{bmatrix}
		\mathbf{\Phi } _{1}^{H}\mathbf{\Phi } _{1}  & \mathbf{\Phi } _{1}^{H}\mathbf{\Phi } _{2} & \cdots &  \mathbf{\Phi } _{1}^{H}\mathbf{\Phi } _{l} \\
	  \mathbf{\Phi } _{2}^{H}\mathbf{\Phi } _{1}  &\mathbf{\Phi } _{2}^{H}\mathbf{\Phi } _{2} & \cdots &  \mathbf{\Phi } _{2}^{H}\mathbf{\Phi } _{l} \\
		\vdots & \vdots & \ddots & \vdots \\
	\mathbf{\Phi } _{l}^{H}\mathbf{\Phi } _{1}  &\mathbf{\Phi } _{l}^{H}\mathbf{\Phi } _{2}  & \cdots & \mathbf{\Phi } _{l}^{H}\mathbf{\Phi } _{l} 
	\end{bmatrix}
	, \
	\mathbf{\bar{u} } =
	\begin{bmatrix}
		\mathbf{u} _{1}  \\ \mathbf{u} _{2}  \\ \vdots \\ \mathbf{u} _{l} 
	\end{bmatrix}
	,
\end{equation}
\begin{equation}
	\mathbf{R}  =
	\begin{bmatrix}
		\mathbf{t } _{1}\mathbf{t } _{1} ^{H} & \cdots & \mathbf{t } _{1}\mathbf{t } _{l} ^{H} \\
		\mathbf{t } _{2}\mathbf{t } _{1} ^{H}  & \cdots & \mathbf{t } _{2}\mathbf{t } _{l} ^{H}\\
		\vdots & \ddots & \vdots \\
		\mathbf{t } _{l}\mathbf{t } _{1} ^{H} & \cdots & \mathbf{t } _{l}\mathbf{t } _{l} ^{H}
	\end{bmatrix}
	,  \
	\mathbf{E}  =
		\begin{bmatrix}
		\mathbf{\Upsilon  } _{1}^{H}\mathbf{\Upsilon } _{1}   & \cdots & \mathbf{\Upsilon  } _{1}^{H}\mathbf{\Upsilon } _{l}  \\
	\mathbf{\Upsilon  } _{2}^{H}\mathbf{\Upsilon } _{1}  & \cdots & \mathbf{\Upsilon  } _{2}^{H}\mathbf{\Upsilon } _{l} \\
		\vdots & \ddots & \vdots \\
	\mathbf{\Upsilon  } _{l}^{H}\mathbf{\Upsilon } _{1}   & \cdots & \mathbf{\Upsilon  } _{l}^{H}\mathbf{\Upsilon } _{l} 
	\end{bmatrix}
	.
\end{equation}
Subsequently, by substituting the extended auxiliary variables into the problem (P3.1), the multi-IRS phase-shift matrices are optimized through DC programming. The FPI method can still be applied to optimize the independent antenna PS, which is alternately optimized with the multi-IRS phase-shift matrices until convergence.

\begin{algorithm}
	\caption{Joint Independent PS Ratio, Receive Beamforming Vector, and Dual-IRS Phase Shifts Optimization Algorithm for Problem (P1)}
	\begin{algorithmic}[1]
		\State Initialize $\left \{ \mathbf{\Theta}  _{a} \right \} ^{\left ( 0 \right ) } $ , $\left \{ \mathbf{\Theta}  _{b} \right \} ^{\left ( 0 \right ) }$, $\left \{ \mathbf{w} \right \} ^{\left ( 0 \right ) } $, $\left \{ \boldsymbol{\rho} \right \} ^{\left ( 0 \right ) }$, convergence threshold $\epsilon$, and iteration index $t=0$.
		\Repeat
		\State For given $\left \{ \mathbf{\Theta}  _{a} \right \} ^{\left ( t \right ) } $ and $\left \{ \mathbf{\Theta}  _{b} \right \} ^{\left ( t \right ) }$, apply MMSE for $\left \{ \mathbf{w} \right \} ^{\left ( t+1 \right ) } $, solve the problem (P2), and obtain independent PS ratio $\left \{ \boldsymbol{\rho} \right \} ^{\left ( t+1 \right ) }$.
		\State For given $\left \{ \mathbf{w} \right \} ^{\left ( t+1 \right ) } $ and $\left \{ \boldsymbol{\rho} \right \} ^{\left ( t+1 \right ) }$, solve the problem (P3.3), and obtain dual-IRS phase shifts $\left \{ \mathbf{\Theta}  _{a} \right \} ^{\left ( t+1 \right ) } $ and $\left \{ \mathbf{\Theta}  _{b} \right \} ^{\left ( t+1 \right ) }$.
		\State Update $t=t+1$.
		\Until{The fractional increase of the objective value is below a threshold $\epsilon$.}
		\State \textbf{return} Independent PS ratio, receive beamforming vector, and dual-IRS phase shifts.
	\end{algorithmic}
\end{algorithm}

\section{SWIPT Optimization and Analysis for the Hybrid-field Model}
\label{section:far}
In the near-field channels, the dual-IRS achieve the optimal passive beamforming gains for the considered SWIPT system with optimal phase shifts, while their performance may differ in the hybrid-field channels. This section is dedicated to dual-IRS-aided SWIPT optimization for the hybrid-field model. Beyond optimizing the independent PS and receive beamforming, we conduct a further analysis of the dual-IRS characteristics to extract valuable insights.

\subsection{SWIPT Optimization}
Define $\mathbf{g}_{a}= \mathbf{G} _{a}\mathbf{\Theta }_{a} \mathbf{h} _{a}$ and $\mathbf{g} _{b}= \mathbf{G} _{b}\mathbf{\Theta }_{b} \mathbf{h} _{b} $ as the combined channels via IRS1 and IRS2, respectively. Furthermore, we define the IRS1-user channel $\mathbf{G} _{a}=\left [ \hat{\mathbf{q} }_{1} ,\hat{\mathbf{q} }_{2},...,\hat{\mathbf{q} }_{M}  \right ]^{T}  \in \mathbb{C }^{M\times {N_a} } $ and the IRS2-user channel $\mathbf{G} _{b}=\left [ \bar{\mathbf{q} } _{1}  ,\bar{\mathbf{q} }_{2},...,\bar{\mathbf{q} }_{M}  \right ]^{T} \in \mathbb{C }^{M\times {N_b} }  $. Since the long distance between the user and the dual-IRS, as well as the rich scattering environment at the side of the user, $\mathbf{G} _{a}$ and $\mathbf{G} _{b}$ are assumed to follow the Rayleigh fading channel model, i.e.,
\begin{align}
\hat{\mathbf{q} }_{m}\sim \mathcal{CN} \left ( 0,\frac{\beta }{d_{a} ^{\alpha } } \boldsymbol{I}   \right ),\ m=1,2,...,{M} ,\\
\quad  \bar{\mathbf{q} }_{m}\sim \mathcal{CN} \left ( 0,\frac{\beta }{d_{b}^{\alpha }  } \boldsymbol{I}   \right ), \ m=1,2,...,{M} ,
\end{align}
where $\alpha $ accounts for the pass loss exponent between the user and dual-IRS, $\beta $ represents the reference path gain at a distance of 1 meter, and $d_{a}$ and $d_{b}$ stand for the distances between the user and dual-IRS. 

For any beamforming vector $\theta_a $($\theta_b$) , $\mathbf{\Theta } _a$($\mathbf{\Theta } _b$) is an independent diagonal unitary matrix that does not influence the distribution of $\hat{\mathbf{q} }_{m}$($\bar{\mathbf{q} }_{m}$). Hence, by invoking the Lindeberg-Levy central limit theorem, we have $\mathbf{g} _{a}\sim \mathcal{CN} \left ( 0,\varrho _{a}^2   \boldsymbol{I}   \right )$ and $\mathbf{g} _{b}\sim \mathcal{CN} \left ( 0,\varrho _{b}^2   \boldsymbol{I}   \right )$, with the average combined channel gains of AP-IRS-user links given by
\begin{align} \label{gain1}
	\varrho _{a}^2&=\sum_{n_{x_a} =-\frac{N_{x_a}-1 }{2}  }^{\frac{N_{x_a}-1 }{2}}\sum_{n_{z_a}=-\frac{N_{z_a}-1 }{2}  }^{\frac{N_{z_a}-1 }{2}} 
	q _{n_{x_a},n_{z_a}} \frac{\beta }{d _{a}^{\alpha }  } \notag \\
	&=\frac{\beta A l_{y_a}  }{4\pi l_{x_a}^3d_{a} ^{\alpha }   }\sum_{n_{x_a} =-\frac{N_{x_a}-1 }{2}  }^{\frac{N_{x_a}-1 }{2}}\sum_{n_{z_a}=-\frac{N_{z_a}-1 }{2}  }^{\frac{N_{z_a}-1 }{2}}   \notag \\
	& \frac{1}{\left ( 1+\bar{r} _{a}^{2} +2n_{x_a}\xi _{a} +\left ( n_{x_a}^2 +n_{z_a}^2 \right ) \xi _{a}^2 \right )^{\frac{3}{2} } }  , 
\end{align}
\begin{align}\label{gain2}
\varrho _{b}^2&=\sum_{n_{x_b} =-\frac{N_{x_b}-1 }{2}  }^{\frac{N_{x_b}-1 }{2}}\sum_{n_{z_b}=-\frac{N_{z_b}-1 }{2}  }^{\frac{N_{z_b}-1 }{2}} 
q _{n_{x_b},n_{z_b}} \frac{\beta }{d _{b}^{\alpha }  } \notag \\
& =\frac{\beta A l_{y_b}  }{4\pi l_{x_b}^3d_{b} ^{\alpha }   }\sum_{n_{x_b} =-\frac{N_{x_b}-1 }{2}  }^{\frac{N_{x_b}-1 }{2}}\sum_{n_{z_b}=-\frac{N_{z_b}-1 }{2}  }^{\frac{N_{z_b}-1 }{2}}   \notag \\
& \frac{1}{\left ( 1+\bar{r} _{b}^{2} +2n_{x_b}\xi _{b} +\left ( n_{x_b}^2 +n_{z_b}^2 \right ) \xi _{b}^2 \right )^{\frac{3}{2} } }  ,
\end{align}
It can be observed that when the user is located in the far field, the combined channel gains are independent of the beamforming vectors of the dual-IRS. 
In addition, by invoking the Lindeberg-Levy central limit theorem again, we have $\mathbf{g} \sim \mathcal{CN} \left ( 0,\left ( \varrho _{a}^2 +\varrho _{b}^2   \right )  \boldsymbol{I}  \right )$, where $\mathbf{g}$ is the total combined channel for the hybrid-field model, i.e., $\mathbf{g} =\mathbf{g} _{a}+\mathbf{g} _{b}$. 
Accordingly, the total harvested power by the EH in the hybrid-field can be further derived by
\begin{equation}
Q=\eta \sum_{m=1}^{M} (1-\rho _{m} )(P_\mathrm{t}  \left ( \varrho _{a}^2 +\varrho _{b}^2   \right )+P_\mathrm{in}\left |f_{m} \right |^{2}+\sigma_{r}^{2} ) .
\end{equation}

Also at ID, MMSE is adopted for the receive beamforming vector, i.e., $\mathbf{w} = \mathbf{S}^{-1} \mathbf{\Lambda }^{\frac{1}{2}} \mathbf{g}\left ( \boldsymbol{l } \right )  $ with $\mathbf{S}  =P_\mathrm{in} \mathbf{\Lambda }^{\frac{1}{2}} \mathbf{f}\mathbf{f}^{H} \mathbf{\Lambda }^{\frac{1}{2}}+\sigma _{r} ^{2}\mathbf{\Lambda }+\delta  ^{2}\mathbf{I}$. And we can obtain the average received SINR in (\ref{sinrb}) as 
\begin{align}
\overline{\mathrm{SINR}}  =\mathbb{E} \left \{ \mathrm{SINR} \right \} =\frac{P_\mathrm{t}\displaystyle\sum_{m=1}^{M}\rho _{m}\left ( \varrho _{b}^2 +\varrho _{b}^2   \right )  }{P_\mathrm{in} \displaystyle\sum_{m=1}^{M}\rho _{m} \left | f_{m}  \right | ^{2}  +\displaystyle\sum_{m=1}^{M}\rho _{m}\sigma _{r}^{2}+ \delta ^{2}}.
\end{align}

Since the average SINR and the combined channel gains are independent of the IRS reflection matrices, the problem (P1) can be transformed into a feasibility-check problem (P4), which is given by
\begin{align}
(\text{P4}) \ \  &  \ \ \underset{  \boldsymbol{\rho}  }{\text{max}} \quad  Q 
\notag      \\
\mbox{s.t.}\quad  
& \ \frac{P_\mathrm{t}\displaystyle\sum_{m=1}^{M}\rho _{m}\left ( \varrho _{b}^2 +\varrho _{b}^2   \right )  }{P_\mathrm{in} \displaystyle\sum_{m=1}^{M}\rho _{m} \left | f_{m}  \right | ^{2}  +\displaystyle\sum_{m=1}^{M}\rho _{m}\sigma _{r}^{2}+ \delta ^{2}}\ge \gamma _{0} , \label{rnk} \\ 
& \ 0\le \rho _{m}\le 1, \ m=1,2,...,M.
\end{align}

Obviously, the problem (P4) is  a standard convex optimization problem, which can be solved by applying the CVX toolbox \cite{grant2016cvx}.

It is worth noting that the analysis and optimization principles presented above can also be extended to multi-IRS systems in the hybrid-field case. In particular, based on the preceding analysis, we can further obtain the total combined channel $\mathbf{g} \sim \mathcal{CN} \left ( 0,\sum_{l=1}^{L}\varrho _{l}^2   \boldsymbol{I}  \right )$, where the average combined channel gains of the $l$-th AP-IRS-user link can be expressed as
\begin{align}  \label{hybrid_mul}
	\varrho _{l}^2=&\frac{\beta A l_{y_l}  }{4\pi l_{x_l}^3d_{l} ^{\alpha }   }\sum_{n_{x_l} =-\frac{N_{x_l}-1 }{2}  }^{\frac{N_{x_l}-1 }{2}}\sum_{n_{z_l}=-\frac{N_{z_l}-1 }{2}  }^{\frac{N_{z_l}-1 }{2}}   \notag \\
	&\quad \quad \quad \frac{1}{\left ( 1+\bar{r} _{l}^{2} +2n_{x_l}\xi _{l} +\left ( n_{x_l}^2 +n_{z_l}^2 \right ) \xi _{l}^2 \right )^{\frac{3}{2} } }  .
\end{align}
Accordingly, substituting (\ref{hybrid_mul}) into the problem (P4) allows the optimal solution of multi-IRS systems to be obtained via CVX.

\subsection{Dual-IRS Characteristics Analysis}

Unlike the dual-IRS-aided SWIPT system in near-field channels, in hybrid-field channels, the system performance is independent of the phase shifts of the dual IRSs. Still, it is determined by the number of reflecting elements, as indicated by equations (\ref{gain1}) and (\ref{gain2}). To further explore the role of dual-IRS for SWIPT when the user is located in the far-field, we derive closed-form expressions for the combined channel gains of the AP-IRS-user links, and then analyze their asymptotic performance.

%Different from dual-IRS in the near-field channels, in the hybrid-field channels, dual-IRS do not require specific phase shift to improve the channel gains according to equations (\ref{gain1}) and (\ref{gain2}).

\begin{lemma}
Under the practical conditions of $\xi _{a}=\varepsilon /l_{x_a}  \ll  1 $ and $\xi _{b}=\varepsilon /l_{x_b}  \ll  1 $, the closed-form expression for the combined channel gain of IRS1 is given by
\begin{align} \label{ra1}
	&\varrho _{a}^2=\frac{\beta A  }{2\pi \varepsilon^{2}    d_{a} ^{\alpha }   } \notag  \\
	&\left ( \arctan \frac{\xi _{a}N_{z_a}\left (1+\xi _{a}N_{x_a}/2  \right )   } {2\bar{r} _{a}\sqrt{N_{x_a}^2\xi _{a}^2/4+N_{z_a}^2\xi _{a}^2/4+N_{x_a}\xi _{a}+\bar{r} _{a}^2+1} } \right. \notag  \\
	&\left. -\arctan \frac{\xi _{a}N_{z_a}\left ( 1-\xi _{a}N_{x_a}/2  \right )   }{2\bar{r} _{a}\sqrt{N_{x_a}^2\xi _{a}^2/4+N_{z_a}^2\xi _{a}^2/4-N_{x_a}\xi _{a}+\bar{r} _{a}^2+1} }  \right ) ,
\end{align}
and the resultant combined channel gain of IRS2 is similarly given by
\begin{align}\label{sa2}
	&\varrho _{b}^2=\frac{\beta A  }{2\pi \varepsilon^{2}    d_{b} ^{\alpha }   } \notag  \\
	&\left ( \arctan \frac{\xi _{b}N_{z_b}\left ( 1+\xi _{b}N_{x_b}/2  \right )   } {2\bar{r} _{b}\sqrt{N_{x_b}^2\xi _{b}^2/4+N_{z_b}^2\xi _{b}^2/4+N_{x_b}\xi _{b}+\bar{r} _{b}^2+1} } \right. \notag  \\
	&\left. -\arctan \frac{\xi _{b}N_{z_b}\left ( 1-\xi _{b}N_{x_b}/2  \right )   }{2\bar{r} _{b}\sqrt{N_{x_b}^2\xi _{b}^2/4+N_{z_b}^2\xi _{b}^2/4-N_{x_b}\xi _{b}+\bar{r} _{b}^2+1} }  \right ) .
\end{align}
\end{lemma}
{\emph {Proof:}}
Please refer to Appendix A for the detailed proof.
$\hfill\blacksquare$

Accordingly, given the element spacing $\varepsilon$ and substituting $\xi _{a}=\varepsilon /l_{x_a} $ and $\xi _{b}=\varepsilon /l_{x_b}$ into (\ref{ra1}) and (\ref{sa2}), respectively, it is found that the average combined channel gains increase with decreasing distances between the dual-IRS and AP along the x-axis. This is expected since the pass loss of the combined channels can be reduced by decreasing the AP-IRS link distances. Therefore, it is preferable to have the near-field IRS-aided AP to achieve higher combined channel gains over the SWIPT system. Besides, it is easily verified that $\varrho _{a}^2$ and $\varrho _{b}^2$ increase monotonically with the number of elements along the x-axis and/or z-axis of the IRS on the respective link. Accordingly, due to the IRS not supplying any passive beamforming gain in the hybrid-field model, the combined channel gain just monotonically increases with the number of IRS elements but does not grow with the square of the number of IRS elements as in the traditional far-field plane wave model.
\begin{proposition}
With IRS1 expanding, the combined channel gain via IRS1 can be further derived as
\begin{equation}\label{rou_a}
\varrho _{a}^2= 
\left\{ 
    \begin{array}{lc}      
        \frac{\beta A  }{\pi \varepsilon ^{2}d_{a} ^{\alpha }  }\arctan \frac{\xi _{a}N_{z_a}   } {2\bar{r} _{a}} \qquad  \qquad \qquad \ , \mathrm{condition \  a} ,\\
        \frac{\beta A  }{2\pi \varepsilon ^{2}d_{a} ^{\alpha }  } \arctan \frac{4\xi _{a}\bar{r} _{a}^2 N_{x_a}  } {4\bar{r} _{a}^2+4-\xi _{a}^2 N_{x_a}^2} \quad \ \ \  \ \ \ , \mathrm{condition \  b} , \\
        \frac{\beta A  }{2\pi \varepsilon ^{2}d_{a} ^{\alpha }  } \left ( \pi +\arctan \frac{4\xi _{a}\bar{r} _{a}^2 N_{x_a}  } {4\bar{r} _{a}^2+4-\xi _{a}^2 N_{x_a}^2}  \right )  , \mathrm{condition \  c} .  \\
%        \frac{\beta A  }{2\varepsilon ^{2}d_{a} ^{\alpha }  }\quad \quad\quad \qquad \qquad \qquad \quad \  \ \ \ \ \ , \mathrm{condition \  d} 
    \end{array}
\right.
\end{equation}
where condition $a$ represents that $N_{x_a}$ is sufficiently large, condition $b$ represents that $N_{z_a} $ is sufficiently large but $N_{x_a}< \frac{2\sqrt{1+\bar{r} _{a}^2} }{\xi _{a}}$, and condition $c$ represents that $ N_{z_a}$ is sufficiently large but $ N_{x_a}>\frac{2\sqrt{1+\bar{r} _{a}^2} }{\xi _{a}}$. The Eq. (\ref{rou_a}) is similar to the combined channel gain via IRS2, i.e., $\varrho _{b}^2$, when IRS2 is expanding.
\end{proposition}
{\emph {Proof:}}
Please refer to Appendix B for the detailed proof.
$\hfill\blacksquare$
 
\begin{figure}
	\centering
	\includegraphics[width=0.55\linewidth]{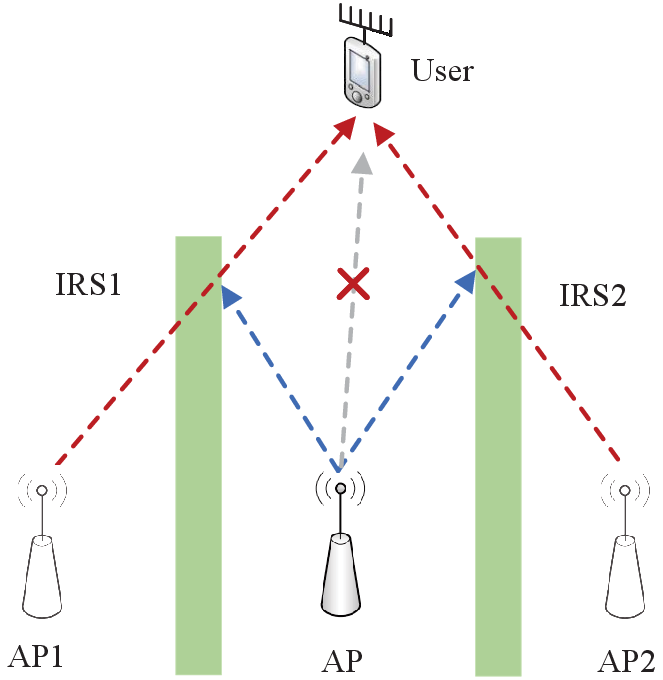}\\
	\caption{\small{SWIPT system with dual-IRS and its equivalence in the hybrid-field model.}}\label{mirro}
	\vspace{-0.2cm}
\end{figure}

It can be observed from {\bf Proposition 2} that when the IRS is expanding along a single coordinate axis direction, the combined channel gain tends to a finite value related to the number of elements in the other coordinate axis direction. On the other hand, when the IRS is expanding simultaneously along both coordinate axis directions, the combined channel gain gradually approaches a constant value of $\frac{\beta A  }{2\varepsilon ^{2}d ^{\alpha }  }$, instead of increasing infinitely. These observations align with physical intuition: each IRS can at most reflect half of the AP's transmit power \cite{zheng2023simu}. Especially in the extreme case where the array fully occupies the aperture, i.e., $\frac{A}{\varepsilon^2} = 1$, the dual IRSs behave like double mirrors (see Fig. \ref{mirro}), each reflecting half of the transmitted power within its forward-facing reflective region.
Thus, the mirror points of AP with respect to the IRS1 and IRS2, i.e., AP1 and AP2 as shown in Fig. \ref{mirro}, can be considered as two APs operating for the SWIPT system. To put it another way, the dual-IRS introduce double transmitters to create high received power and SINR conditions for the user in this case.

\section{Simulation Results}
In this section, simulation results are provided to examine the performance of the dual-IRS-aided SWIPT system with independent PS in both the near- and hybrid-field models. Under the 3D Cartesian coordinate system shown in Fig. \ref{3d}, the AP equipped with a single isotropic antenna is placed at the origin. The IRS1 and IRS2 are placed on the x-z plane with their centers at (1m,1m,0m) and (1m,-1m,0m), respectively. The location of IN is set as (100m,100m,0m). We consider that the user is equipped with $M$ = 5 antennas and the antenna spacing is $d$ = 0.2 m. In the near-field case, the location of the user's central antenna is randomly distributed within a circle of radius 1m centered at (30m, 0m, -2m), while in the hybrid-field scenario, the circle is centered at (50m, 0m, -2m). Unless otherwise stated, the wavelength is $\lambda_0$ = 0.4 m, the IRS element spacing is set as  $\varepsilon=\lambda_0 /2 $ = 0.2 m, and the physical size of each reflecting element is set as $A=\varepsilon^{2}$ that is consistent with \cite{zheng2023simu}. We assume that both IRSs have the same number of reflecting elements, i.e., $N_0=N_a=N_b$, and for each IRS, we set $N_0=N_xN_z$ where $N_x$ and $N_z$ denote the number of reflecting elements along the x-axis and z-axis, respectively, and we fix $N_x=N_z=11$. We set $\sigma _{r}^{2}=\delta  ^{2}= -50 \mathrm{dBm}$ and $\eta=0.9$ in our numerical simulations. For the channel from IN to the user, i.e., $\mathbf{f} $, we set the path loss to $-30 \mathrm{dB}$ when the reference distance is 1m and the path loss exponent to 2.

\subsection{Near-field SWIPT performance}
\begin{figure}
	\centering
	\setlength{\abovecaptionskip}{-0.1cm}
	\includegraphics[width=7cm]{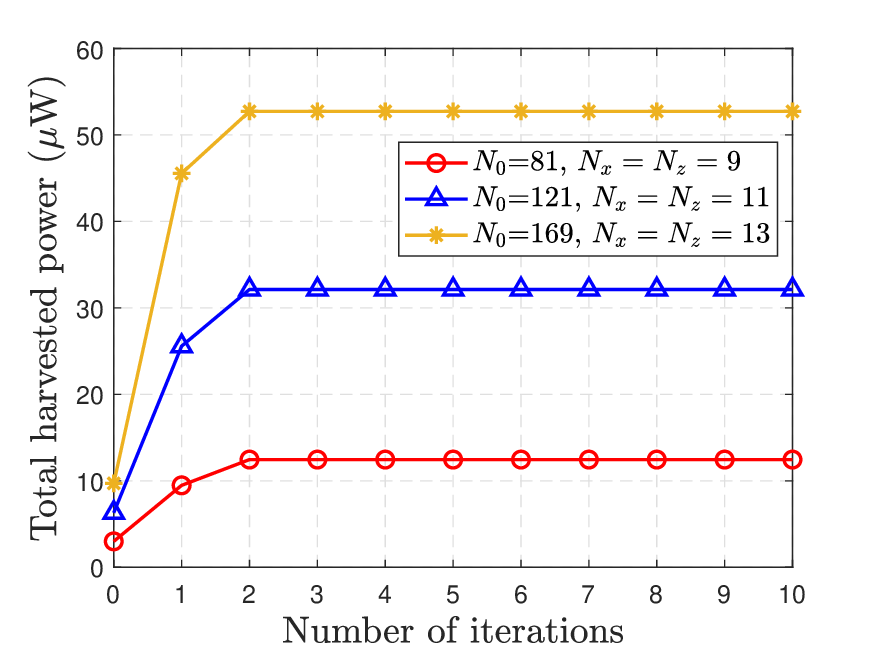}\\
	\caption{\small{The convergence of the proposed algorithm.}}\label{iteration}
	\vspace{-0.1cm}
\end{figure}

First, we focus on the performance of the considered dual-IRS-aided interference-limited SWIPT system for the near-field model. Unless otherwise stated, the transmit powers of AP and IN are $P_\mathrm{t} =1\mathrm{W}$ and $P_{\mathrm{in} }=2\mathrm{W}$, respectively. The convergence threshold of the proposed algorithm is set as $10^{-5}$. We evaluate the convergence behavior of the proposed algorithm. Fig. \ref{iteration} illustrates the variation in the total harvested power with respect to the number of iterations under different numbers of IRS1/IRS2 reflecting elements. It can be observed that the total harvested power increases with the number of iterations. The proposed algorithm converges at the second iteration, demonstrating its excellent convergence performance. Specifically, we compare the performance of the proposed algorithm with the total number of IRS1/IRS2 reflecting elements set to 81, 121, and 169, respectively. The results indicate that a larger number of IRS1/IRS2 reflecting elements leads to higher total harvested power, which also verifies the effectiveness of the IRS aided SWIPT system.

For comparison, we consider the following baseline schemes: (1) FPI+SDR: Semidefinite relaxation (SDR) is utilized to optimize the dual-IRS phase shifts (see \cite{wu2019netw}), while FPI is retained to obtain the optimal independent PS ratio. (2) IPM+DC: The independent PS ratio is optimally obtained by applying the interior-point method (see \cite{lis2016indep}), while DC programming is employed to optimize the dual-IRS phase shifts. (3) Com-algorithm: The algorithm is the same as the optimization of the proposed algorithm for the two sub-problems, except that the two sub-problems are not optimized alternately. (4) EAPS+OPS: Alternating optimization of traditional equal PS and dual-IRS phase shifts (see \cite{li2022beam}). (5) IAPS+RPS: Each antenna at the receiver adopts optimal independent PS factors, wheres fixing phase shifts at dual IRSs.

\begin{figure}
	\centering
	\setlength{\abovecaptionskip}{-0.1cm}
	\includegraphics[width=7cm]{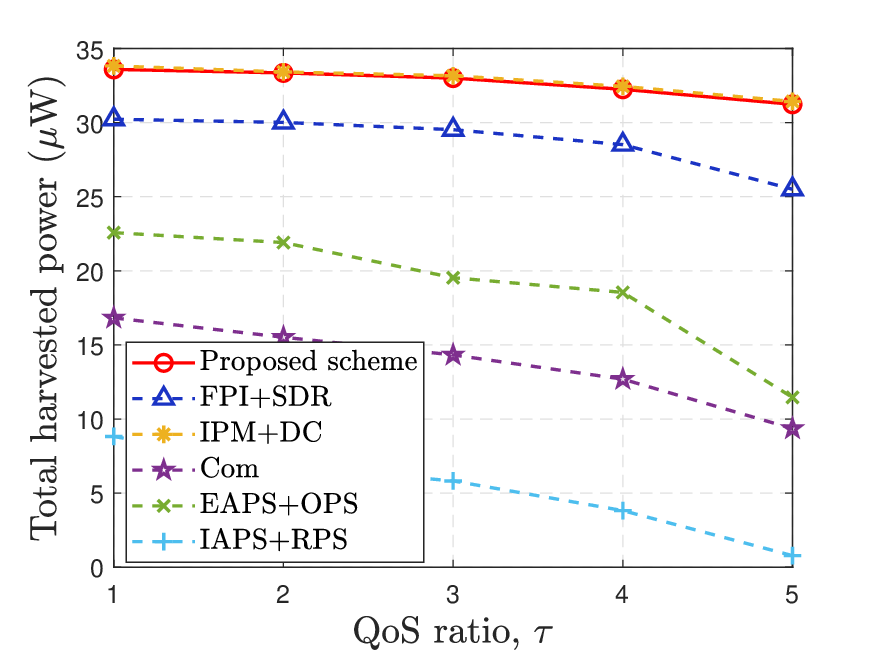}\\
	\caption{\small{The total harvested power versus QoS ratio in the near-field model.}}\label{Qos_E}
	\vspace{-0.3cm}
\end{figure}

\begin{figure}
	\centering
	\setlength{\abovecaptionskip}{-0.1cm}
	\includegraphics[width=7cm]{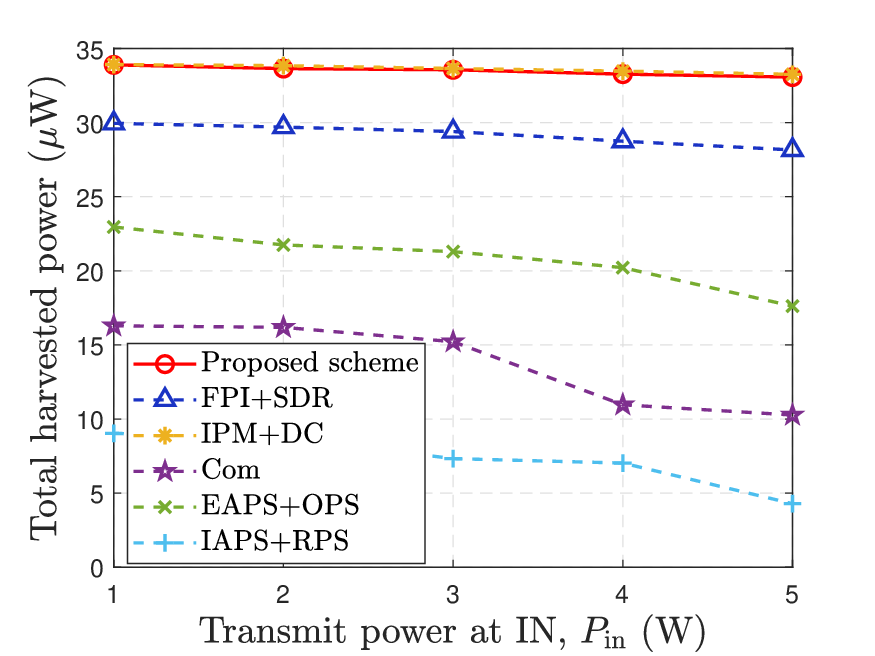}\\
	\caption{\small{The total harvested power versus transmit power at IN in the near-field model.}}\label{q_E}
	\vspace{-0.4cm}
\end{figure}

For ease of exposition, we take $\tau$ as the QoS ratio and set $\gamma _{0}=\tau  \bar{\gamma } _{0} $ (in linear scale) with $\bar{\gamma } _{0} =10$. In Fig. \ref{Qos_E}, we plot the total harvested power by the EH versus the QoS ratio in the near-field model. It can be seen that the total harvested power under different schemes decreases with the increase of the QoS ratio. This is because the higher the QoS threshold, the larger the requirement for information decoding, resulting in fewer resources allocated to the EH. By leveraging dual-IRS passive beamforming to strengthen the desired signal and independent PS to dynamically allocate received power, the proposed system balances the dual effects of interference.
Additionally, the proposed scheme and the IPM+DC scheme yield nearly identical harvested power, with the latter being slightly higher. This negligible performance gap indicates that the proposed algorithm achieves near-optimal results.
Compared with the IPM that requires solving convex subproblems with a complexity of $\mathcal{ O} \left (M^3  \right ) $, the proposed FPI algorithm involves simple closed-form updates with a per-iteration complexity of $\mathcal{ O} \left (M^2  \right ) $, resulting in significantly lower computational cost and improved scalability, while incurring a slight performance loss.
Furthermore, our proposed scheme achieves an average performance improvement of 14.7\% over the FPI+SDR scheme. The reason is that the SDR relaxes the rank-one constraint and relies on randomization, causing performance loss. In contrast, DC iteratively linearizes the non-convex term while preserving each subproblem convex, which can be efficiently solved via CVX. Although the DC method involves iterative updates with a total complexity of $\mathcal{ O} \left (T_{DC} N_0^6  \right ) $, it achieves better accuracy and faster convergence than the single-shot SDR with complexity $\mathcal{ O} \left (N_0^6  \right ) $.
In addition, the results show that adopting independent PS significantly expands the QoS-power region compared to   optimal traditional equal PS. 
Compared to the com-algorithm, the superior performance of our proposed algorithm is mainly because the latter is considered from the perspective of global optimization, while the former only performs local optimization. 
The proposed scheme significantly outperforms the IAPS+RPS scheme, as the fixed phase shifts of the dual IRSs are unable to adapt to the channel conditions, reducing the design freedom of the system and consequently lowering the received signal power.

\begin{figure}
	\centering
	\setlength{\abovecaptionskip}{-0.05cm}
	\includegraphics[width=7cm]{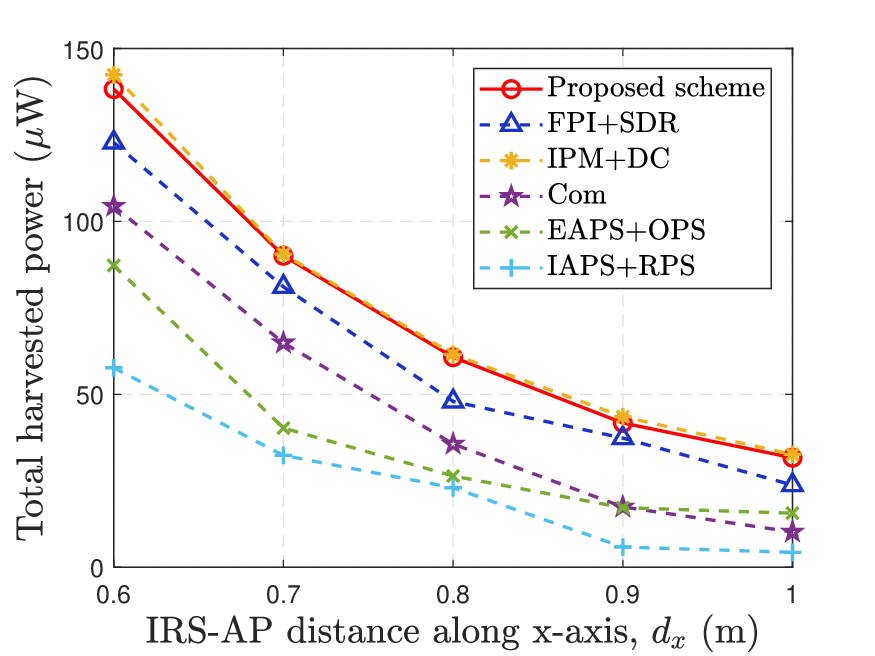}\\
	\caption{\small {The total harvested power versus IRS-AP distance along x-axis in the near-field model.}}\label{d_E}
	\vspace{-0.4cm}
\end{figure}

Fig. \ref{q_E} elaborates how the total harvested power varies with the transmit power at IN in the near-field model with QoS ratio $\tau=1$. In general, as the transmit power at IN increases, the total harvested power continues to decrease. This is mainly because the detrimental impact of stronger interference on information decoding outweighs its limited contribution to energy harvesting under the SINR requirement. It is found that setting independent PS factors on different antennas continuously increases the harvested power compared to the optimal traditional equal PS. When the transmit power at IN is 5W, the harvested power of the proposed scheme is 17.5\% higher than that of the EAPS+OPS scheme. This can be attributed to the fact that the flexible per-antenna adaptation provided by independent PS enables efficient power harvesting from interference while maintaining reliable information decoding. 
Moreover, the proposed scheme outperforms the IAPS+RPS scheme, demonstrating that the dual-IRS effectively strengthens the desired signal through intelligent reflection and, in conjunction with the adaptive PS design, improves SINR robustness while simultaneously facilitating the harvesting of useful energy from interference for the user.
In comparison to the FPI+SDR and Com algorithms, the proposed algorithm achieves superior performance, demonstrating its effectiveness and advantages.
%Compared with the random phase shift and random PS cases, the proposed scheme can significantly increase the total harvested power, because these two random schemes do not optimize the passive beamforming and receiver architecture, respectively, and the random phase and PS may even degrade system performance. The reason why the proposed algorithm performs better than the com-algorithm is that the latter does not achieve the convergence of the entire problem. 
In Fig. \ref{d_E}, we depict the total harvested power versus the IRS1/IRS2-AP distance along the x-axis, i.e., $d_x=l_{x_a}=l_{x_b}$, with $P_{\mathrm{in} }=2\mathrm{W}$ and $\tau=1$. As we can see in Fig. \ref{d_E}, the total harvested power under different schemes increases with the decrease of distance between IRS1/IRS2 and AP, which thus indicates that positioning the dual-IRS in proximity to the AP can substantially augment the performance of the SWIPT system, as observed in \cite{wu2021intelligent}. This observation aligns with the intuition that placing the IRS near the AP results in stronger incident signal power and a more effective reflecting aperture, where the non-uniform illumination enables a spatial focusing effect, enhancing the equivalent channel power gain and overall system performance. In addition, when the IRS1/IRS2-AP distance along the x-axis remains the same, the proposed scheme closely approaches the IPM+DC scheme, yet delivers a significant performance improvement compared to the other baseline schemes.

\subsection{Hybrid-field SWIPT performance}

\begin{figure}
	\centering
	\setlength{\abovecaptionskip}{-0.1cm}
	\includegraphics[width=7cm]{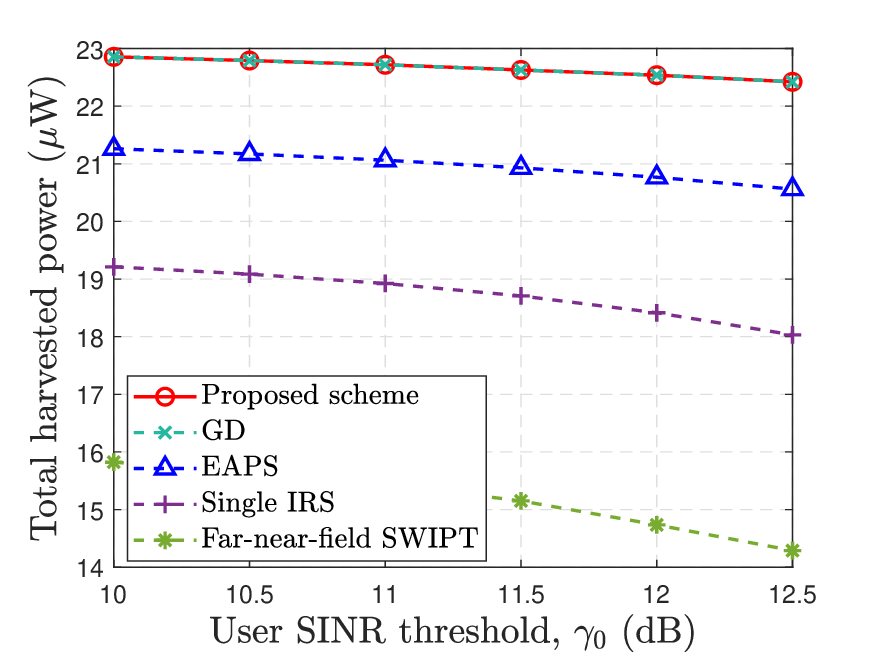}\\
	\caption{\small{The total harvested power versus SINR threshold in the hybrid-field model.}}\label{hyfield_SINR}
	\vspace{-0.2cm}
\end{figure}

\begin{figure}
	\centering
	\setlength{\abovecaptionskip}{-0.1cm}
	\includegraphics[width=7cm]{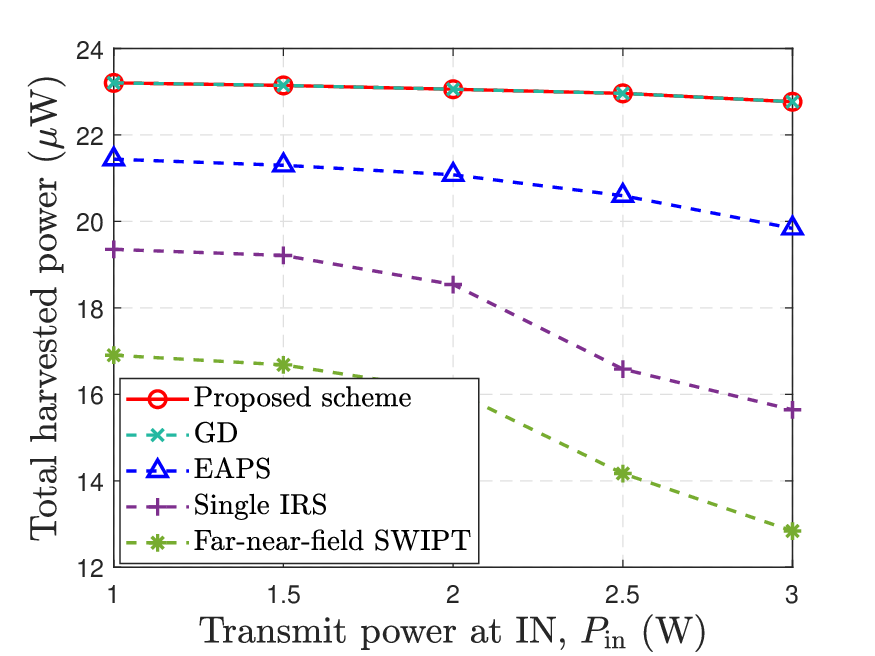}\\
	\caption{\small{The total harvested power versus transmit power at IN in the hybrid-field model.}}\label{hyfield_q}
	\vspace{-0.4cm}
\end{figure}

Next, we concentrate on the performance of the considered dual-IRS-aided interference-limited SWIPT system for the hybrid-field model. We set the pass loss exponent to $\alpha=1.6$ and the reference path gain at a distance of 1m to $\beta=-30\mathrm{dB}$ for the channel between the dual-IRS and the user. Unless specified otherwise, the transmit powers of AP and IN are $P_\mathrm{t} =10\mathrm{W}$ and $P_{\mathrm{in} }=2\mathrm{W}$, respectively. Since the dual-IRS do not require specific phase shifts for SWIPT optimization in the hybrid-field channels, the performance of SWIPT mainly depends on the size of the dual-IRS (i.e., the number of elements), as well as the independent PS factors in this case.
Therefore, we consider the following baselines to verify the effectiveness of our proposed scheme in improving SWIPT performance for the hybrid-field case: (1) GD: Employing gradient descent method for the optimization of the problem (P4). (2) EAPS: Applying optimal equal antenna PS to the user antenna array. (3) Single IRS: A single IRS with optimal independent PS factors is mounted with the same total number of elements as the dual IRSs, i.e., $2N_0$, shares its central position with IRS1, and features an extended size along the x-dimension compared to IRS1. (4) Far-near-field SWIPT: The central positions of IRS1 and IRS2 are located at (48m,1m,0m) and (48m,-1m,0m), respectively, while keeping the locations of the AP and the user unchanged. Thus, the AP–IRS link operates in the far-field, while the IRS–user link is in the near-field.

\begin{figure}
	\centering
	\setlength{\abovecaptionskip}{-0.1cm}
	\includegraphics[width=7cm]{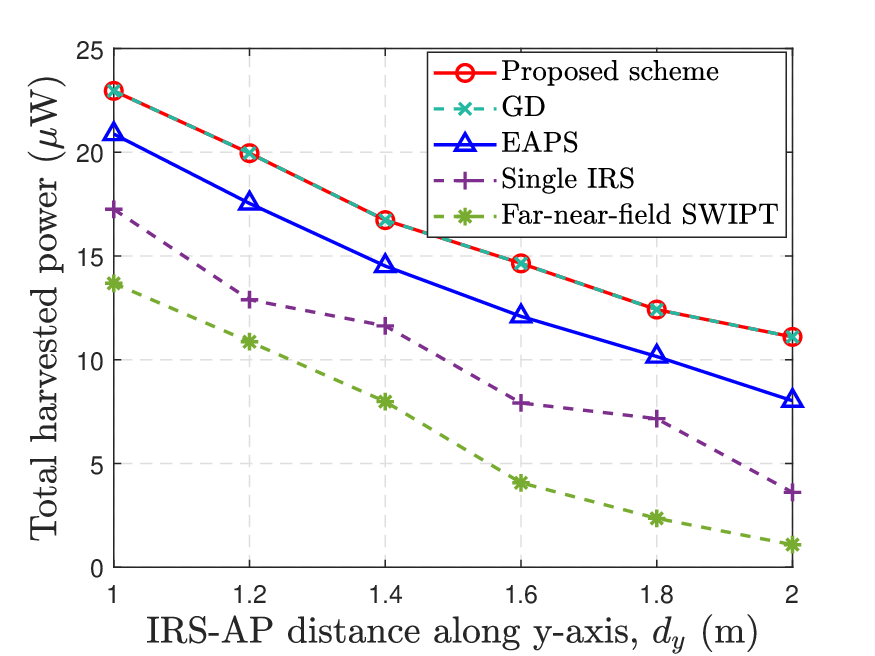}\\
	\caption{\small{The total harvested power versus IRS-AP distance along y-axis in the hybrid-field model.}}\label{hybrid_d}
	\vspace{-0.5cm}
\end{figure}

In Fig. \ref{hyfield_SINR}, we show the total harvested power by the EH versus the user's SINR threshold in the hybrid-field model. Several interesting observations are made as follows. First, it can be observed that the total harvested power under different schemes decreases with the increase of the SINR threshold. This is intuitively expected for the reason that a greater proportion of the intended signal needs to be reserved for information decoding to maintain a high SINR threshold, resulting in reduced energy supply for EH. Second, the performance curves obtained applying CVX and the manually implemented gradient descent method almost overlap, confirming that both the correctness of our convex reformulation and the numerical stability of the adopted solver. Since CVX internally employs efficient interior-point solvers and provides a convenient modeling interface for convex programming, it ensures both implementation simplicity and numerical robustness, making it suitable for validating the proposed optimization scheme. Third, the average harvested power of our proposed scheme surpasses that of the equal PS scheme by 2 $\mu$W, representing an 7.5\% increase, which verifies that independent PS can also provide a flexible trade-off between power harvesting and information decoding in the hybrid-field case. Moreover, our proposed scheme achieves a significant performance improvement compared to the the single IRS scheme with independent PS applied at the receiver. This is because the dual IRS in the proposed hybrid-field case can be viewed as double mirrors, where the AP is projected by the dual IRSs as two APs operating in the SWIPT system. Conversely, a single IRS functions merely as a solitary reflective surface, leading to a substantial attenuation in transmit power and, as a result, a markedly lower total harvested energy compared to the dual-IRS scheme. In addition, all schemes in the proposed hybrid-field case of near-filed AP–IRS link and far-filed IRS user link outperform the far-near-field case, with the largest gap of 8.1 $\mu$W in total harvested power at an SINR of 12.5 dB observed for the proposed dual-IRS scheme with independent PS. This phenomenon lies in that placing the IRS near the user increases the AP–IRS distance, weakening the reflected dual-IRS signal due to severe path loss. In contrast, with a near-field AP–IRS link, the IRS can spatially focus the reflected signal and receive a stronger incident wave, enabling more efficient energy reflection to the user.

\begin{figure}
	\centering
	\setlength{\abovecaptionskip}{-0.1cm}
	\includegraphics[width=7cm]{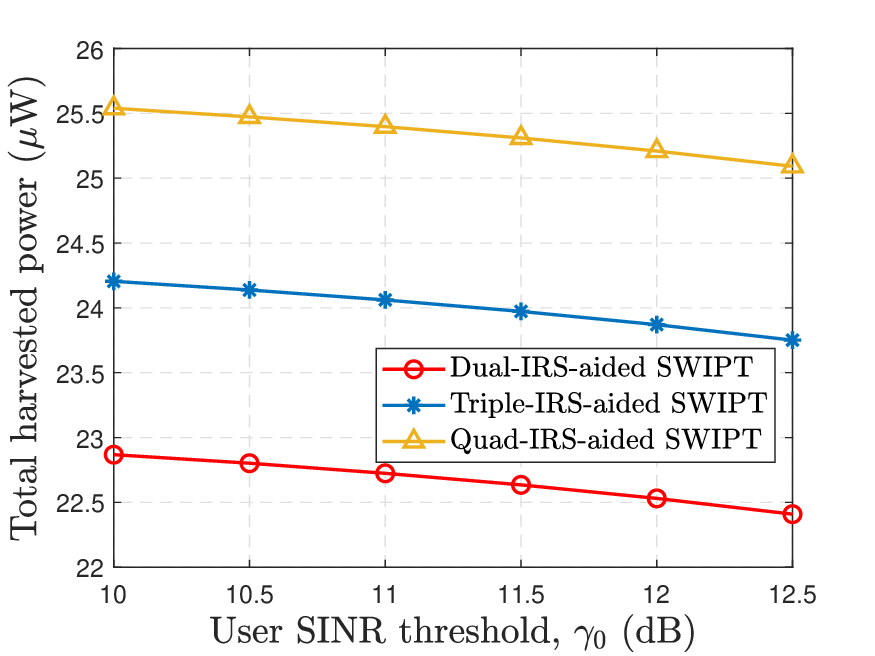}\\
	\caption{\small{Comparison of multi-IRS systems.}}\label{multi_IRS}
	\vspace{-0.25cm}
\end{figure}

In Fig. \ref{hyfield_q}, we investigate the total harvested power versus the transmit power at IN with $\gamma _{0} = 10\mathrm{dB}$ in the hybrid-field model. First, it can be seen that the performance of the SWIPT system with independent PS is better than that of the system with equal PS. Furthermore, when the transmit power at IN increases, the gap of this performance becomes larger. 
The underlying reason is that, unlike the EAPS scheme where all receiving antennas share an optimal identical PS ratio, the proposed independent PS scheme enables each antenna to adapt its PS ratio according to the interference it experiences, allowing a fine-grained trade-off between energy harvesting and information decoding, which becomes particularly advantageous under interference-enhanced conditions.
Meanwhile, our proposed scheme achieves a significant performance improvement compared to the single IRS and the far-near-field SWIPT scheme. This stems from the fact that placing the dual-IRS near the AP allows more effective reflection and focusing of the transmitted signal, which boosts the received signal power, enhances the effective SINR, and facilitates meeting the users' QoS requirements, thereby contributing to the interference tolerance.
In Fig. \ref{hybrid_d}, we plot the total harvested power versus the IRS1/IRS2-AP distance along the y-axis, i.e., $d_y=l_{y_a}=-l_{y_b}$. It is observable that the harvested power of all the IRS-aided SWIPT schemes in both our considered hybrid-field case and the far-near-field baseline increase as the IRS1/IRS2-AP distance along the y-axis decreases, which is attributed to the increased signal reflection power in the IRS-reflected link. 
Moreover, the proposed scheme almost overlaps with the GD benchmark, which verifies the effectiveness of the optimization method.

Fig. \ref{multi_IRS} compares the total harvested power versus the SINR threshold for different multi-distributed-IRS aided SWIPT systems, where each IRS is equipped with $N_0$ reflecting elements, with the centers of the third and fourth IRS arrays positioned at (3m,1m,0m) and (3m,-1m,0m), respectively.
It can be observed that the total harvested power improves with the increasing number of IRSs (i.e., dual-, triple-, and quadruple-IRS configurations). This is attributed to the additional reflecting surfaces and propagation paths introduced by multiple IRSs, which strengthen the total combined channel and improve energy harvesting performance.
From a system design perspective, although multi-IRS architectures offer certain performance gains, they also introduce increased system complexity, control overhead, and deployment cost. In contrast, the dual-IRS configuration strikes a desirable balance between performance enhancement, implementation complexity, and deployment cost, thus serving as a representative and practical architecture for evaluating multi-IRS-aided SWIPT systems.

\begin{figure}
	\centering
	\setlength{\abovecaptionskip}{-0.1cm}
	\includegraphics[width=7cm]{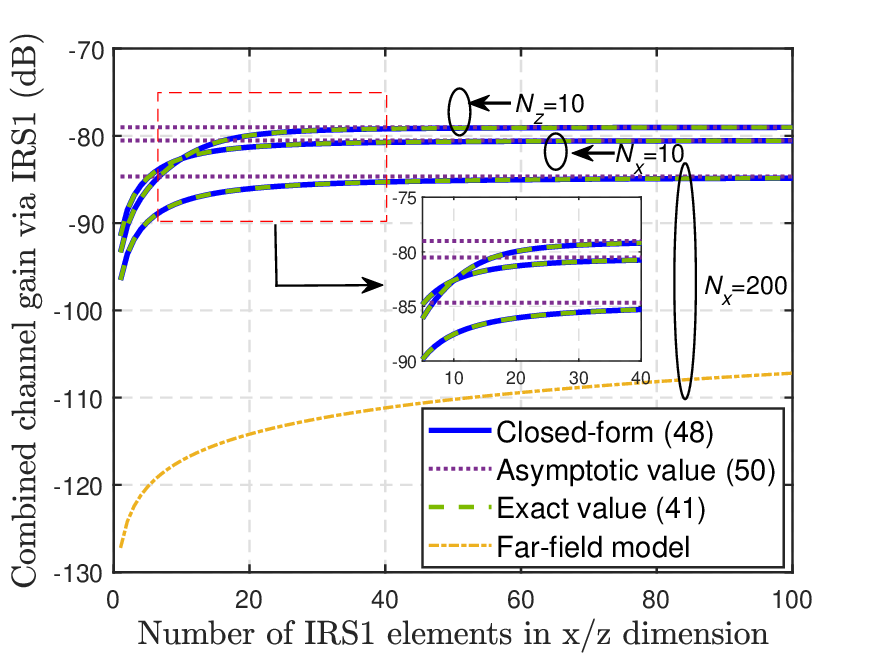}\\
	\caption{\small{The combined channel gain via IRS1 versus the number of IRS1 elements in x/z dimension.}}\label{N_gain}
	\vspace{-0.25cm}
\end{figure}

Finally, we show the combined channel gain $\varrho _{a}^2$ for the AP-IRS1-user link versus the number of IRS1 elements along the x/z-axis in Fig. \ref{N_gain}. For the sake of comparison, we consider the far-field  uniform-plane wave propagation model for the AP-IRS1 channel by substituting $q  _{ n_{x_a},n_{z_a} }=\frac{A l_{y_a}}{4\pi l_{x_a}^{3} \left ( 1+\bar{r}_a^2  \right )^{\frac{3}{2} }   }$ to (\ref{gain1}), while the IRS1-user link still follows the Rayleigh fading channel model. Additionally, the exact value of $\varrho _{a}^2$ derived by (\ref{gain1}), the closed-form expression calculated in (\ref{ra1}), and the asymptotic gain limits under the three conditions given in (\ref{rou_a}) are also shown in the Fig. \ref{N_gain}. Owing to the symmetric distribution of IRS1 and IRS2 relative to the AP, the curve shown in Fig. \ref{N_gain} also applies to IRS2. For simplicity, we set $N_x=N_{x_a} $, $N_z=N_{z_a}$, and $N=N_xN_y$. A few noteworthy observations can be highlighted. It is first seen that whether IRS1 extends along the x or z dimension, the closed-form expression proposed in {\bf Lemma 1} aligns seamlessly with the exact values. Moreover, the cases depicted in Fig. \ref{N_gain}, where $N_z$ = 10, $N_x$ = 10, and $N_x$ = 200, correspond to the three distinct conditions outlined in {\bf Proposition 2}. As illustrated, when the number of elements in the x/z dimension surpasses a certain threshold, the exact values converge asymptotically to the limits predicted by {\bf Proposition 2}, affirming the validity and precision of the derived asymptotic expressions in {\bf Proposition 2}.
Besides, it can be observed that as the total number of reflecting elements increases, the combined channel gain of the hybrid-field model initially scales linearly with $N$, and gradually approaching saturation when $N_x$ exceeds 40, while remaining higher than that obtained under the far-field model, whose combined channel gain increases approximately linearly with $N$. This is because when the IRS moves from the near-field to the far-field region, the reflection characteristics of the IRS evolve from spatially focused to directional reflection. The former achieves stronger performance by concentrating energy within a spatial region, whereas the latter relies on gradual energy accumulation for linear gain.

\section{Conclusion}
In this paper, we investigated a new dual-IRS aided SWIPT system with independent antenna PS, and optimized system performance in the near- and hybrid-field models. Specifically, we established an optimization problem to maximize the harvested power by jointly optimizing the independent PS ratio, receive beamforming vector, and dual-IRS phase shifts in both near-field and hybrid-field cases. For the near-field model, the maximization problem was decomposed into two sub-problems to address its non-convexity, and the optimization variables were alternately optimized through the Lagrange duality method and the DC programming algorithm. For the hybrid-field model, the non-convex problem can be transformed into a solvable one by exploiting the fact that the combined channel gains are independent of the dual-IRS reflection matrices.  
Additionally, we analyzed the characteristics of the dual-IRS and derived closed-form expressions for the combined channel gains, which demonstrate that when the IRS surfaces are sufficiently large, the dual-IRS can be regarded as double mirrors, effectively forming two virtual APs serving the user.
Numerical results demonstrated that our proposed scheme can significantly augment the harvested power compared to other baseline schemes in both channel models.
%\textcolor{blue}{In the future, it is an interesting direction to consider the multi-antenna AP by jointly designing the multi-antenna AP’s active beamforming, IRS’s passive beamforming, and independent antenna PS.}  
%highlighting the critical auxiliary role of dual-IRS in the AP's near-field and the importance of applying independent PS to each receiving antenna, which can substantially alleviate the pressure on the AP with low cost in practice.

{\appendices

	\section{Proof of the Lemma 1}
Under the practical condition of $\xi _{a}=\varepsilon /l_{x_a} \ll  1$, we approximate the double summation of (\ref{gain1}) with the double integral as
\begin{align} \label{lemma2rou}
	\varrho _{a}^2&\!=\frac{\beta A l_{y_a}  }{4\pi l_{x_a}^3d_{a} ^{\alpha }   }\sum_{n_{x_a} =-\frac{N_{x_a}-1 }{2}  }^{\frac{N_{x_a}-1 }{2}}\sum_{n_{z_a}=-\frac{N_{z_a}-1 }{2}  }^{\frac{N_{z_a}-1 }{2}}   \notag \\
	&\quad  \frac{1}{\left ( 1+\bar{r} _{a}^{2} +2n_{x_a}\xi _{a} +\left ( n_{x_a}^2 +n_{z_a}^2 \right ) \xi _{a}^2 \right )^{\frac{3}{2} } }  \notag \\
	&\triangleq \frac{\beta A l_{y_a}  }{4\pi l_{x_a}^3d_{a} ^{\alpha }   }\frac{1}{\xi_a ^2}\int_{\!-\!\frac{\xi_a\! N_{x_a} }{2} }^{\frac{\xi_{a}\! N_{x_a} }{2}}\int_{\!-\!\frac{\xi_{a} \!N_{z_a} }{2} }^{\frac{\xi_a\! N_{z_a} }{2}} 
	\frac{\mathrm{d} x\mathrm{d} z}{\left ( 1\!+\!\bar{r_a}^{2} \!+\!2x\!+\!x^{2}\!+\!z^2  \right )^{\frac{3}{2} }  }  \notag  \\
	&\triangleq \frac{\beta A \bar{r}_a  }{4\pi \varepsilon ^{2}  d_{a} ^{\alpha }   }\underbrace{ \int_{-\!\frac{\xi_a N_{x_a} }{2} }^{\frac{\xi_a N_{x_a} }{2}}\int_{-\!\frac{\xi_a N_{z_a} }{2} }^{\frac{\xi_a N_{z_a} }{2}}
		\frac{\mathrm{d} x\mathrm{d} z}{\left ( 1\!+\!\bar{r}_a ^{2}\! +\!2x\!+\!x^{2}\!+\!z^2  \right )^{\frac{3}{2} }  }  }_I.
\end{align}
By performing the integration with respect to $z$ followed by $x$, the double integral in (\ref{lemma2rou}) can be further derived as
\begin{align} \label{lem2close}
	I&\!=\!\int_{-\frac{\xi_a N_{x_a} }{2} }^{\frac{\xi_a N_{x_a} }{2}} 
	\frac{z}{1\!+\!\bar{r_a} ^{2}\!+\!2x\!+\!x^{2}} \!\!\left. \frac{\mathrm{d} x}{\left ( 1\!+\!\bar{r}_a ^{2}\! +\!2x\!+\!x^{2}\!+\!z^2  \right )^{\frac{1}{2} }  }\right|_{-\frac{\xi_a N_{z_a} }{2}}^{\frac{\xi_a N_{z_a} }{2}} \notag  \\
	=& \!\!\int_{-\frac{\xi_a N_{x_a} }{2} }^{\frac{\xi_a N_{x_a} }{2}}\frac{\xi_a N_{z_a}\mathrm{d} x }{\left ( 1\!+\bar{r}_a ^{2}+2x+x^{2}  \right ) \left ( 1\!+\!\bar{r}_a ^{2} +2x+x^{2}+\!N_{z_a}^2 \xi_a ^{2} /4  \right )^{\frac{1}{2} }} \notag  \\
	=&\!\!\left.  \frac{2 \xi_a N_{z_a} }{\sqrt{N_{z_a}^2 \xi_a ^{2}  \bar{r}_a ^{2}   } } \! \arctan \! \left ( \frac{\xi_a N_{z_a}\left ( 1+x \right ) }{2\bar{r}_a\!\sqrt{1\!+\!\bar{r}_a ^{2}\! +\!2x \!+ \! x^{2}\!+\!N_{z_a}^2\! \xi_a ^{2} /4} }   \right ) \right|_{\!-\!\frac{\xi_a \! N_{x_a} }{2}}^{\frac{\xi_a N_{x_a} }{2}} \notag  \\ 
	=& \frac{2}{\bar{r}_a} \left ( \arctan \frac{\xi_a N_{z_a}\left ( 1+\xi_a N_{x_a}/2  \right )   } {2\bar{r}_a \sqrt{N_{x_a}^2\xi_a ^2/4+N_{z_a}^2\xi_a ^2/4+N_{x_a}\xi_a +\bar{r}_a^2+1} } \right. \notag  \\
	&\left. -\arctan \frac{\xi_a N_{z_a}\left ( 1-\xi_a N_{x_a}/2  \right )   }{2\bar{r}_a \sqrt{N_{x_a}^2\xi_a^2/4+N_{z_a}^2\xi ^2/4-N_{x_a}\xi +\bar{r}_a ^2+1} }  \right ) .
\end{align}
By substituting (\ref{lem2close}) into (\ref{lemma2rou}), we obtain the expression shown in (\ref{ra1}). The proof process for (\ref{sa2}) is similar to that for $\varrho _{a}^2$, which is ignored for brevity. According to all of the above, the proof is completed.

	\section{Proof of the Proposition 2}
With the condition that $N_{x_a}$ is sufficiently large, we have $\xi _{a}N_{x_a}/2\gg 1 $ and $N_{x_a}^2\xi _{a}^2/4 \gg N_{z_a}^2\xi _{a}^2/4+N_{x_a}\xi _{a}+\bar{r} _{a}^2+1$, so that Eq. (\ref{ra1}) can be calculated as 
\begin{align} 
	 \varrho _{a}^2 = &\frac{\beta A  }{2\pi \varepsilon^{2}    d_{a} ^{\alpha }   } \! \left ( \! \arctan \frac{\xi _{a}^2N_{z_a}/2  } {\bar{r} _{a} \xi _{a}} \! - \! \arctan \frac{-\xi _{a}^2N_{z_a}/2  } {\bar{r} _{a} \xi _{a}} \! \right  )\notag  \\
	& = \frac{\beta A  }{\pi \varepsilon ^{2}d_{a} ^{\alpha }  }\arctan \frac{\xi _{a}N_{z_a}   } {2\bar{r} _{a}} . 
\end{align}
Furthermore, with the condition $N_{z_a}$ is large enough, we have $N_{z_a}^2\xi _{a}^2/4 \gg N_{x_a}^2\xi _{a}^2/4+N_{x_a}\xi _{a}+\bar{r} _{a}^2+1$, and then
\begin{align} \label{ra2}
	\varrho _{a}^2&= \frac{\beta A  }{2\pi \varepsilon^{2}    d_{a} ^{\alpha }   } \notag  \\
	&\left ( \arctan \frac{\left (1+\xi _{a}N_{x_a}/2  \right )   } {\bar{r} _{a} }  - \arctan \frac{\left (1-\xi _{a}N_{x_a}/2  \right )   } {\bar{r} _{a} } \right ) .
\end{align}
Let $x_1 =  \frac{\left (1+\xi _{a}N_{x_a}/2  \right )   } {\bar{r} _{a} }$ and $x_2 =  \frac{\xi _{a}\left (1-N_{x_a}/2  \right )   } {\bar{r} _{a} } $. When $x_1x_2>-1$, i.e., $ N_{x_a}< \frac{2\sqrt{1+\bar{r} _{a}^2} }{\xi _{a}}$, the equation (\ref{ra2}) can be simplified as 
\begin{align} 
	\varrho _{a}^2=  \frac{\beta A  }{2\pi \varepsilon ^{2}d_{a} ^{\alpha }  } \arctan \frac{4\xi _{a}\bar{r} _{a}^2 N_{x_a}  } {4\bar{r} _{a}^2+4-\xi _{a}^2 N_{x_a}^2}.
\end{align}
On the other hand, when $x_1x_2<-1$, i.e., $ N_{x_a}> \frac{2\sqrt{1+\bar{r} _{a}^2} }{\xi _{a}}$, the combined channel gain is approximated as 
\begin{align} 
	\varrho _{a}^2=  \frac{\beta A  }{2\pi \varepsilon ^{2}d_{a} ^{\alpha }  } \left ( \pi +\arctan \frac{4\xi _{a}\bar{r} _{a}^2 N_{x_a}  } {4\bar{r} _{a}^2+4-\xi _{a}^2 N_{x_a}^2}  \right ).
\end{align}
%Finally, let $y_1=\frac{\xi _{a}N_{z_a}\left (1+\xi _{a}N_{x_a}/2  \right )   } {2\bar{r} _{a}\sqrt{N_{x_a}^2\xi _{a}^2/4+N_{z_a}^2\xi _{a}^2/4+N_{x_a}\xi _{a}+\bar{r} _{a}^2+1} }$ and $y_2 = \frac{\xi _{a}N_{z_a}\left ( 1-\xi _{a}N_{x_a}/2  \right )   }{2\bar{r} _{a}\sqrt{N_{x_a}^2\xi _{a}^2/4+N_{z_a}^2\xi _{a}^2/4-N_{x_a}\xi _{a}+\bar{r} _{a}^2+1} } $ . With the conditions $N_{z_a} \to\infty $ and $  N_{x_a} \to\infty$, we have $y_1\to\infty$ and  $y_2\to-\infty$. As such, the equation (\ref{ra}) can be reduced as
%\begin{align} 
%	\lim_{N_{z_a} \to \infty } \varrho _{a}^2&=  \frac{\beta A  }{2\pi \varepsilon ^{2}d_{a} ^{\alpha }  } \left ( \frac{\pi}{2}-(-\frac{\pi}{2} )   \right ) =\frac{\beta A  }{2\varepsilon ^{2}d_{a} ^{\alpha }  }.
%\end{align}
The proof process for $\varrho _{b}^2$ is similar to that for $\varrho _{a}^2$, which is ignored for brevity. According to all of the above, the proof is completed.

}

\balance

\end{document}